\documentclass[letterpaper,11pt]{article}
\usepackage[utf8]{inputenc}

\usepackage{jheppub}
\usepackage{graphicx,verbatim}
\usepackage{blindtext}
\usepackage[T1]{fontenc}
\usepackage{epsf}
\usepackage{lmodern}
\usepackage[mathscr]{euscript}

\usepackage{ytableau}
\usepackage{youngtab}

\usepackage{amsmath}
\usepackage{amsfonts}
\usepackage{amssymb}
\usepackage{mathrsfs}
\usepackage{slashed}
\usepackage{xcolor}
\usepackage{tikz, float}
\usetikzlibrary{patterns,shapes.misc}

\def\ri{\mathrm{i}}

\def\bT{\mathbf{T}}
\def\bt{\mathbf{t}}

\def\BC{\mathbb{C}}
\def\BM{\mathbb{M}}

\def\BZ{\mathbb{Z}}

\def\CalN{\mathcal{N}}

\def\CalR{\mathcal{R}}

\def\CalS{\mathcal{S}}
\def\CalC{\mathcal{C}}

\def\Tr{{\rm Tr}}

\def\tb{\mathtt{b}}

\def\tp{\mathtt{p}}
\def\tq{\mathtt{q}}

\def\tw{\mathtt{w}}

\def\fg{\mathfrak{g}}

\def\EH{\EuScript{H}}

\def\ET{\EuScript{T}}

 \def\p{\partial}
 
 \def\a{\alpha}
 \def\b{\beta}
 \def\g{\gamma}
 \def\d{\delta}

  \def\sgn{\text{sgn}}

\def\beq{\begin{equation}}
\def\eeq{\end{equation}}

\newtheorem{thm}{Theorem}[section]
\newtheorem{defn}[thm]{Definition}

\title{Dimers for Type D Relativistic Toda Model}
\author[a]{Kimyeong Lee}  
\author[b]{and Norton Lee}
\affiliation[a]{School of Physics, Korea Institute for Advanced Study, \\Hoegiro 85, Seoul 02455, Korea}
\affiliation[b]{Center for Geometry and Physics, Institute for Basic Science (IBS), \\Pohang 37673, Korea}
\emailAdd{klee@kias.re.kr, norton.lee@ibs.re.kr}

\preprint{KIAS-P24038, CGP24008}

\vspace{2cm}
\abstract{We construct dimer graphs for type D relativistic Toda models by introducing impurities into the $Y^{2N,0}$ square dimer graphs. By properly placing the impurities and changing the canonical variables assigned to the 1-loops on the dimer graph, we perform a "folding" of the graphs, which yields the type D relativistic Toda lattice Hamiltonian and monodromy matrix. }

\begin{document}

\maketitle


\newpage

\section{Introduction}

The relativistic Toda lattice (RTL) was first introduced in \cite{ruijsenaars1990relativistic} with a given Lie algebra $\mathfrak{g}$. There are two Lax formalisms for RTL: the Lax triad, which consists of three $N \times N$ Lax matrices \cite{ragnisco1989periodic,bruschi1989lax}, and Sklyanin's $2 \times 2$ $R$-matrix formalism \cite{sklyanin1988boundary,sklyanin1995separation}.

Type-A RTL, which is defined based on the Lie algebra $\fg = A_{N-1}$, belongs to a class of cluster integrable systems proposed by Goncharov and Kenyon \cite{Kenyon2011dimers}: an integrable system corresponding to a periodic planar dimer placed on a torus \cite{Eager:2011dp}. The dual graph of the dimer graph is a planar, periodic quiver. The quiver gauge theory, called $Y^{p,q}$, arises from a stack of D3 branes probing a single toric Calabi-Yau three-fold (CY3). Both the quiver and the dimer can be constructed based on the toric diagram of the CY3 \cite{Hanany:2005ve}. Interestingly, recent studies show that there exist general dimer graphs on top of the "standard" dimer built from the $Y^{p,q}$ quiver \cite{Lee:2023wbf}. These dimer graphs share the same toric diagram but have different conserving Hamiltonians.

A natural question to ask is whether the correspondence between the dimer and RTL can be extended to RTLs defined on Lie algebras other than type A. In this note, we will focus on the type D RTL defined on the affine Lie algebra $\hat{D}_N$. Our goal is to find the correct dimer graph whose perfect matching reproduces the commuting Hamiltonians of the type D RTL.

It turns out that the connection between relativistic integrable systems and five-dimensional $\mathcal{N}=1$ supersymmetric gauge theory proves to be quite useful in the search for the correct dimer graph. The connection between supersymmetric gauge theories and algebraic integrable models, known as the \emph{Bethe/Gauge correspondence}, was first observed between four-dimensional $\mathcal{N}=2$ supersymmetric gauge theories and non-relativistic integrable systems. The geometry of the low-energy states of the four-dimensional $\mathcal{N}=2$ supersymmetric gauge theory is identified with the phase space of an algebraic integrable system \cite{Gorsky:1999gx,GorskyGM_1_N=2,GorskyM_Manybody,Donagi:1995cf}. The Seiberg-Witten curve of the supersymmetric gauge theory coincides with the spectral curve of the algebraic integrable model, thus sharing the same toric diagram.

The Bethe/Gauge correspondence can be extended to a supersymmetric gauge theory with gauge group $G$ and eight supercharges, compactified on a circle of radius $r$. Its Seiberg-Witten curve corresponds to the spectral curve of the relativistic integrable system with $\mathfrak{g}=\text{Lie}(G)$ algebra. The radius $r = \frac{1}{c}$ acts as the inverse of the speed of light. Indeed, the Bethe/Gauge correspondence is how the Seiberg-Witten curve for five-dimensional $\mathcal{N}=1$ theory was constructed \cite{Nekrasov:1996cz}.

In this note, we will focus on finding the correct dimer graph for the $\hat{D}_N$ RTL. The Bethe/Gauge correspondence provides an advantage, as the corresponding gauge theory can be embedded into a larger $SU$ gauge theory and then reduced in degrees of freedom by folding. In the case of type D, $SO(2N)$ SYM shares the same toric diagram with $SU(2N)+8F$ gauge theory with eight fundamental hypermultiplets fine-tuned \cite{Hayashi:2023boy}. The O5-plane in the brane construction of the $SO(2N)$ gauge group indicates that a proper folding from $SU(2N)+8F$ is required to obtain the correct degrees of freedom. This also restricts how the dimer graph should be constructed.

In this note, we construct the dimer graph based on the type A square dimer with impurities or flavors introduced. By properly placing the impurities and identifying parameters through folding, we successfully construct the dimer graph for the $\hat{D}_N$ RTL. The monodromy matrix of the type D RTL is constructed from the Kasteleyn matrix of the dimer, thus recovering all the conserving Hamiltonians.

\section*{Outine}

This paper is organized as follows: 
In Section.~\ref{sec:RTL} we give a quick review on the type D RTL and its integrability in the Skyanin's Lax formalism. 
In Section.~\ref{sec:Dimer} we review the construction of square dimer model for type A RTL. 
We intoduce impurity reflecting the modification based on the fundamental matter in the toric diagram $SU(2N)+8F$. 
We construct the dimer graph that give rise to the correct first $\hat{D}_N$ RTL Hamiltonain after proper folding and change of canonical variables. 
In Section.~\ref{sec:Lax} we recover Lax/monodromy matrix of $\hat{D}_{N}$ RTL based on the Kasteleyn matrix of the dimer we constructed.  

Finally we point out our conclusion and furture direction in Section.~\ref{sec:summary}

\acknowledgments

The authors thank Saebyeok Jeong, Hee-cheol Kim, Minsung Kim, Yongchao Lu, Xin Wang for useful discussion and correspondence. 
The research of NL is supported by IBS project IBS-R003-D1. KL is supported in part by KIAS Grants
PG006904 and the National Research Foundation of Korea (NRF) Grant funded by the Korea
government (MSIT) (No. 2017R1D1A1B06034369). 
KL also thanks KITP for the program \href{https://www.kitp.ucsb.edu/activities/strings24}{What is
String Theory? Weaving Perspectives Together}. 
This research was supported in part by grant NSF
PHY-2309135 to the Kavli Institute for Theoretical Physics (KITP).

\section{Classical Relativistic Toda of type D}\label{sec:RTL}


The relativistic Toda lattice describes $N$ particles on a line or a ring. The $n$-th particle's position and momentum are denoted by $\tq_n$ and $\tp_n$, which satisfy the Poisson commutation relation:
\begin{align}\label{eq:commu-q-p}
    \{\tq_n,\tp_m\} = \d_{nm}, \ n,m=1,\dots,N.
\end{align}

For a given Lie algebra $\fg$, the relativistic Toda lattice (RTL for short) is defined on its simple root system $\CalR_\fg$ \cite{ruijsenaars1990relativistic,bogoyavlensky1976perturbations}. In this paper, we focus on the Toda lattice defined on the affine Lie algebra of type D. The Hamiltonian of the relativistic Toda lattice associated with the root system of the $\hat{D}_N$ Lie algebra \cite{kuznetsov1992infinite,Kuznetsov:1994ur} is given by:
\begin{align}\label{def:H-RTL-D}
    {\rm H}_{\hat{D}_N} = {\rm H}_0 + {\rm J}_1 + {\rm J}_N
\end{align}
where ${\rm H}_0$ is the open relativistic Toda lattice Hamiltonian of type $A$:
\begin{align}
    {\rm H}_0 = \sum_{n=1}^{N} 2\cosh(\tp_n) + 2{g}^2 \sum_{n=1}^{N-1} e^{\tq_{n}-\tq_{n+1}} \cosh\frac{\tp_{n}+\tp_{n+1}}{2},
\end{align}
and
\begin{align}
\begin{split}
    {\rm J}_1 & = 2 {g}^2 e^{-\tq_1-\tq_2} \cosh \frac{\tp_1-\tp_2}{2} + {g}^4  e^{-2\tq_2},
    \\
    {\rm J}_N & = 2 {g}^2 e^{\tq_N+\tq_{N-1}} \cosh \frac{\tp_N-\tp_{N-1}}{2} + {g}^4 e^{2\tq_{N-1}}.
\end{split}
\end{align}
Here, ${g}$ is the coupling constant.
The total number of terms (written individually in exponential form) in the $\hat{D}N$ RTL Hamiltonian \eqref{def:H-RTL-D} is $2N + 2(N-1) + 3 + 3 = 4N + 4$. Note that there are two additional terms, ${g}^4 e^{-2\tq_2}$ and ${g}^4 e^{2\tq{N-1}}$, which do not originate from the simple roots of $\hat{D}_N$.

In the non-relativistic limit, we scale the coupling constant $g \to rg$ and the momentum $\tp \to r \tp$, where $r$ is the inverse of the speed of light. The relativistic Hamiltonian can be expanded in powers of $r$ as follows:

\begin{align} 
{\rm H}_{\hat{D}_N} & = 2N + 2r^2 \left[ \sum_{n=1}^N \frac{\tp_n^2}{2} + g^2 \sum_{n=1}^{N-1} e^{\tq_n - \tq_{n+1}} + g^2 e^{-\tq_1 - \tq_2} + g^2 e^{\tq_N + \tq_{N-1}} \right] + \mathcal{O}(r^4). 
\end{align}

The coefficient of $r^2$ in this expansion corresponds precisely to the Hamiltonian of the type D non-relativistic Toda lattice.

\subsection{Integrability}

The integrability of relativistic Toda lattice (RTL) is characterized by the $R$-matrix $R_{a_i,a_j} :V_{a_i} \otimes V_{a_j} \to V_{a_i} \otimes V_{a_j}$, which satisfies the Yang-Baxter equation:
\begin{align}
    R_{a_1,a_2}(x-x') R_{a_1,a_3}(x) R_{a_2,a_3}(x') = R_{a_2,a_3}(x') R_{a_1,a_3}(x) R_{a_1,a_2}(x-x').
\end{align}
The $2 \times 2$ Lax matrix is a special case of this $R$-matrix with the choice of:
\[
    V_{a_1}=V_{a_2} = \BC^2 := V_{\rm aux}, \ V_{a_3} = \EH_n
\]
defined for the $n$-th particle.
$\EH_n$ is the Hilbert space of a particle, and $V_{\rm aux} = \BC^2$ is called the auxiliary space. 
On each of the lattice site we define a $2 \times 2$ Lax operator as a $GL_2$-valued function  \cite{Kuznetsov:1994ur,Iorgov:2007ks,deVega:1993xi} 
\begin{align}
    L_n(x) = \begin{pmatrix}
        2\sinh \frac{x - \tp_n}{2} & - {g} e^{-\tq_n} \\
          {g} e^{\tq_n} & 0
    \end{pmatrix}\in \text{End}(\EH_n \otimes V_\text{aux})
\end{align}
where $\tp_n=-\hbar \p_{\tq_n}$ and $\tq_n$ are the canonically conjugated momentum and coordinate of the $n$-th particle. 
The $R$-matrix is given by
\begin{align}\label{def:R-matrix}
    R_{a_1,a_2}(x-x') = 
    \begin{pmatrix} 
        \sinh\frac{x-x'+\hbar}{2} & 0 & 0 & 0 \\
        0 & \sinh \frac{x-x'}{2} & \sinh\frac{\hbar}{2} & 0 \\
        0 & \sinh\frac{\hbar}{2} & \sinh\frac{x-x'}{2} & 0 \\
        0 & 0 & 0 & \sinh\frac{x-x'+\hbar}{2}
    \end{pmatrix} \in \text{End}(V_\text{aux} \otimes V_\text{aux}).
\end{align}

The commutation relations between two matrix components of the Lax operator is governed by the Yang-Baxter RLL-relation (train track relation)
\begin{align}\label{eq:RLL}
    R_{a_1,a_2}(x-x')L_{a_1}(x) L_{a_2}(x') = L_{a_2}(x') L_{a_1}(x) R_{a_1,a_2}(x-x')
\end{align}
which can be verified true by direct computation.  


The monodromy matrix $\bT(x)$ of the type A relativistic Toda lattice (with periodic boundary conditions) is an ordered product of the Lax matrices across $N$ particles:

\begin{align} 
\bT(x) = L_N(x) L_{N-1}(x) \cdots L_2(x) L_1(x) \in \text{End} \left(\bigotimes_{n=1}^N \mathcal{E} \otimes V_{\text{aux}} \right). 
\end{align}

It is evident that the monodromy matrix $\bT(x)$ satisfies the same Yang-Baxter equation as the Lax operator:
\begin{align}\label{eq:RTT}
    R_{a_1,a_2}(x-x')\bT_{a_1}(x) \bT_{a_2}(x') = \bT_{a_2}(x') \bT_{a_1}(x) R_{a_1,a_2}(x-x')
\end{align}

The spectral curve of the integrable system is defined by introducing the spectral parameter $Y$:

\begin{align} 
\text{q-det} (\bT(x) - Y) = Y^2 - \text{Tr} (\bT(x)) Y + \text{q-det} (\bT(x)) = 0. 
\end{align}


\subsection{RTL with boundary}

E. Sklyanin points out that the monodromy matrix for type BCD relativistic Toda lattices (RTLs) can be obtained by introducing a reflection matrix as a boundary condition \cite{sklyanin1988boundary,kuznetsov1992infinite}. The transfer matrix of an RTL with boundary conditions is given by:
\begin{align}
    \bT(x) = \Tr K_+(x) \bt(x) K_-(x) \bt^{-1}(-x) 
\end{align}
where $\bt(x) \in \text{End} \left(\bigotimes_{n=1}^N \mathcal{E} \otimes V_{\text{aux}} \right)$ is the monodromy matrix of the type A RTL, and $K_\pm(x) \in \text{End}(V_{\text{aux}})$ are the reflection matrices obeying the reflection equations:
\begin{align}
\begin{split}
    & R_{12}(x-x') K_{-,1}(x) R_{21}(x+x'-\hbar) K_{-,2}(x') \\ 
    & = K_{-,2}(x') R_{12}(x+x'-\hbar) K_{-,1}(x) R_{21}(x-x')
\end{split}
\end{align}
and 
\begin{align}
\begin{split}
    & R_{12}(-x+x') K^{T}_{-,1}(x) R_{21}(-x-x'-\hbar) K^{T}_{-,2}(x') \\
    & = K^{T}_{+,2}(x') R_{12}(-x-x'-\hbar) K^{T}_{+,1}(x) R_{21}(-x+x')
\end{split}
\end{align}
Given a simple solution $K_\pm(x)$ to the reflection equation, one can verify that:
\begin{subequations}
\begin{align}
    U^{T}_+(x) & =  \bt_+^T(x) K_+^T(x) (\bt^{-1}_+(-x))^T  \\
    U_-(x) & = \bt_-(x) K_-(x) \bt_-^{-1}(-x)  
\end{align}
\end{subequations}
satisfy the same reflection equation if $\bT_\pm(x)$ satisfy the Yang-Baxter equation \eqref{eq:RTT}. 
The transfer matrix (trace of monodromy matrix) is the generating function of the integral of motion:
\begin{align}\label{def:generating}
\begin{split}
    T(x) & = \Tr \ U_+ U_- = \Tr K_+(x) \bt(x) K_-(x) \bt(-x)^{-1} \\
    & = e^{\frac{N}{2}x} \left[ 1 + \sum_{n=1}^N H_n e^{-nx}  \right]
\end{split}
\end{align}

The reflection matrices $K_\pm$ that satisfy the reflection equation are given by:
\begin{subequations}
\begin{align}
    K_+ (\tq_1,\tp_1) & = \begin{pmatrix}
        \alpha_1^+ e^{\frac{x}{2}} - \alpha_2^+ e^{-\frac{x}{2}} & \d^+(e^x + e^{-x}) - \beta^+ \\
        \g^+ - \d^+(e^x + e^{-x}) & \alpha_2^+ e^{\frac{x}{2}} - \alpha_1^+ e^{-\frac{x}{2}}
    \end{pmatrix} \\
    K_- (\tq_N,\tp_N) & = \begin{pmatrix}
        - \alpha_1^- e^{\frac{x}{2}} + \alpha_2^- e^{-\frac{x}{2}} & \g^- - \d^-(e^x + e^{-x}) \\
        \d^- (e^x + e^{-x} ) - \b^- & -\a_2^- e^{\frac{x}{2}} + \a_1^- e^{-\frac{x}{2}}
    \end{pmatrix}
\end{align}
\end{subequations}
These reflection matrices lead to the following (first) Hamiltonian:
\begin{align}
\begin{split}
    H_{1} = & \sum_{j=2}^{N-1} 2\cosh \tp_j + \sum_{j=2}^{N-2} {g}^2 e^{\tq_j-\tq_{j+1}} 2\cosh \frac{\tp_j+\tp_{j+1}}{2} \\
    & + \b^+ + \b^- + {g} \alpha_1^+ e^{-\frac{\tp_2}{2}-\tq_2} + {g} \alpha_2^+ e^{\frac{\tp_2}{2}-\tq_2} + {g} \alpha_1^- e^{-\frac{\tp_{N-1}}{2}+\tq_{N-1}} + {g} \a_2^- e^{\frac{\tp_{N-1}}{2}+\tq_{N-1}} \\
    & + \d^+ {g}^2  e^{-2\tq_2} + \d^- {g}^2 e^{2\tq_{N-1}}. \\
\end{split}
\end{align}

The boundary reflection matrices $K_\pm$ for the RTL of type D, as discussed in \cite{Kuznetsov:1994ur}, are given by:
\begin{subequations}\label{eq:twistK-D}
\begin{align}
     K_+ & = 
    \begin{pmatrix}
        {g} [e^{\frac{x}{2}} 2\cosh(\tp_1/2-\tq_1) - e^{-\frac{x}{2}} 2\cosh(\tp_1/2+\tq_1)]  & 2\cosh x  - 2\cosh {\tp_1} \\
        {g}^2 [2 \cosh 2\tq_1 - 2 \cosh {x}] & {g} [e^{-\frac{x}{2}} \cosh(\tp_1/2-\tq_1)  -e^{\frac{x}{2}} \cosh(\tp_1/2+\tq_1) ]
    \end{pmatrix}, \\
    K_- & = 
    \begin{pmatrix}
        {g} [e^{-\frac{x}{2}}\cosh(\tp_N/2-\tq_N)-e^{\frac{x}{2}} \cosh(\tp_N/2+\tq_N)]  & {g}^2[2\cosh 2\tq_N - \cosh x]  \\ 
        2\cosh x - 2\cosh {\tp_{N}} & {g}^2 [e^{\frac{x}{2}} \cosh(\tp_N/2-\tq_N) - e^{-\frac{x}{2}} \cosh(\tp_N/2+\tq_N)]
    \end{pmatrix}.
\end{align}
\end{subequations}
The monodromy matrix $\bT(x)$ of $\hat{D}_N$ RTL is defined by incorporating both the reflection matrices and the Lax operators. It is given by:
\begin{align}
    \bT(x) = K_+(x) \tilde{L}_2(x)\cdots \tilde{L}_{N-1}(x) K_-(x) L_{N-1}(x) \cdots L_2(x). 
\end{align}
where the $\tilde{L}_n(x)$ operators are defined as: 
\begin{align}
    \tilde{L}_n(x) = \begin{pmatrix}
        0 & -{g} e^{-\tq_n} \\
         {g} e^{\tq_n} & 2\sinh \frac{x + \tp_n}{2}
    \end{pmatrix} = g^4 L_n^{-1}(-x).
\end{align}
These $\tilde{L}_n(x)$ operators are related to the original Lax operators $L_n(x)$ by
\begin{align} 
    \tilde{L}_n(x) = g^4 L_n^{-1}(-x). 
\end{align}
The construction of $\bT(x)$ involves alternating beteween the reflection matrices $K_+(x)$ and $K_-(x)$ with the Lax operators and their inverses, reflecting the boundary conditions and the periodic structure of the $\hat{D}_N$ RTL. 

The spectral curve of classical integrable system is defined by the characteristic polynomial of the monodromy matrix $\bT(x)$. For RTL, this is given by \footnote{The determinant is replaced by q-determinant when quantizing the integrable system.}
\begin{align}
    0 = \det(\bT(x) - Y) = Y^2 - \Tr \bT(x) Y + \det \bT(x). 
\end{align}
To find the spectral curve, we first compute the determinant of the monodromy matrix 
$\bT(x)$. It is given by the product of the determinants of its individual building blocks:
\[
    \det \bT(x) = {g}^{4N-8} \left(X - X^{-1} \right)^4, \ X = e^x. 
\]
Next, we scale $Y \to Y {g} ^{2N-4} (X-X^{-1})^2$ to simplify the expression
\begin{align}\label{eq:spectral-curve}
    {g} ^{2N-4} (X-X^{-1})^2 Y  - T(X) + {g} ^{2N-4} \frac{(X-X^{-1})}{Y} = 0.
\end{align}
where $T(X)$ denotes the trace term related to $\bT(X)$.
This equation represents the spectral curve associated with the relativistic Toda lattice in the given parameterization.

\begin{figure}[h]
    \centering
    \begin{tikzpicture}[scale=0.6, every node/.style={scale=1.2}]
 \begin{scope}
    \draw[dashed] (-7, 0) -- (7,0);
    \draw (5,0) -- (3,1) -- (2,2) -- (2,3) -- (3,4) -- (5,5) ;
    \draw (-5,0) -- (-3,1) -- (-2,2) -- (-2,3) -- (-3,4) -- (-5,5) ;
    \draw (5,0) -- (3,-1) -- (2,-2) -- (2,-3) -- (3,-4) -- (5,-5) ;
    \draw (-5,0) -- (-3,-1) -- (-2,-2) -- (-2,-3) -- (-3,-4) -- (-5,-5) ;
    \draw (-3,1) -- (3,1);
    \draw (-3,-1) -- (3,-1);
    \draw (2,2) -- (-2,2);
    \draw (2,-2) -- (-2,-2);
    \draw (2,2) -- (-2,2);
    \draw (2,-2) -- (-2,-2);
    \draw (2,3) -- (-2,3);
    \draw (2,-3) -- (-2,-3);
    \draw (3,4) -- (-3,4);
    \draw (3,-4) -- (-3,-4);

    \node[above] at (6,0) {O5$^+$};
    \node[above] at (0,0) {O5$^-$};
    \node[above] at (-6,0) {O5$^+$};
\end{scope}
 \begin{scope}[xshift = 12cm]
 \draw (0, 1.2 * 4) -- (1.2, 1.2 * 2) -- (1.2, - 1.2 * 2) -- (0, - 1.2 * 4) -- (-1.2, - 1.2 * 2) -- (- 1.2, 1.2 * 2)-- (0, 1.2 * 4);
\foreach \i in {-4, -3, ..., 3, 4}
{ \filldraw[] (0, 1.2 * \i) circle (5pt);
}
\foreach \i in {-1, 0, 1}
{ \filldraw[fill=white] (-1.2, 1.2 * \i) circle (5pt);
}
\foreach \i in {-1, 0, 1}
{ \filldraw[fill=white] (1.2, 1.2 * \i) circle (5pt);
}
\filldraw[] (-1.2, 1.2 * 2) circle (5pt);
\filldraw[] (-1.2, -1.2 * 2) circle (5pt);
\filldraw[] (1.2, -1.2 * 2) circle (5pt);
\filldraw[] (1.2, 1.2 * 2) circle (5pt);
\end{scope}
    \end{tikzpicture}
    \caption{The brane construction of $SO(8)$ gauge group with toric nodes indicated by the brane construction. Notice that O$5^+$ branes are replaced by 3 nodes instead of 1. See \cite{Hayashi:2023boy} for detail.}
    \label{fig:so(8) - web}
\end{figure}
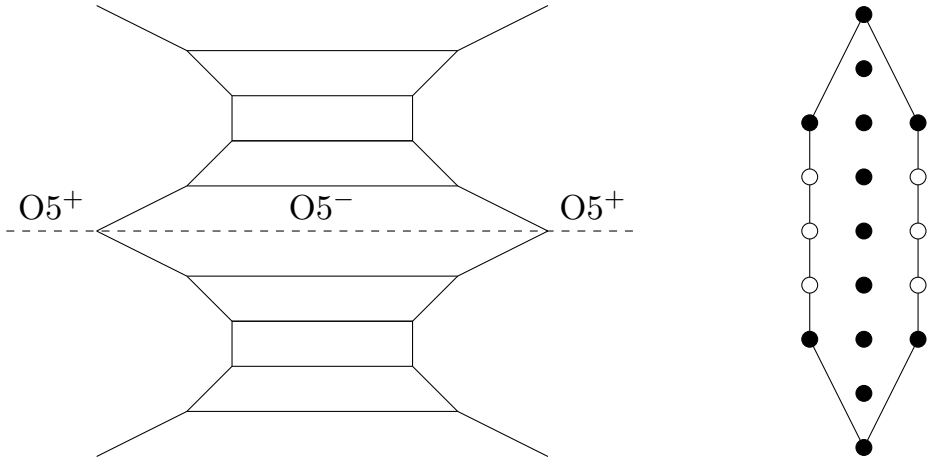

The Type D relativistic Toda provides insights to the description of Seiberg-Witten ansatz for $SO(2N)$ pure gauge theory \cite{Martinec:1995by}. 
For 5d $\CalN=1$ $SO(2N)$ pure super Yang-Mills (SYM) theory, the Seiberg-Witten curve is given by
\begin{align}
    (X-X^{-1})^2 Y + 2 \prod_{\alpha=1}^N (XU_\alpha - X^{-1}U_{\alpha}^{-1}) (XU_\alpha^{-1} - X^{-1}U_{\alpha}) + \frac{(X-X^{-1})^2}{Y} =0
\end{align}
To visualize the corresponding toric diagram $\ET$, one can refer to the D-brane and O-plane construction. The toric diagram encapsulates the geometric data relevant for the gauge theory. For an illustration of this toric diagram in the context of $SO(8)$, see Fig.~\ref{fig:so(8) - web}. 

\section{Dimer integrable system} \label{sec:Dimer}


The relativistic Toda lattice system is part of the broader class of cluster integrable models, which were explored in depth by Kenyon and others \cite{Kenyon2011dimers}. These models can be constructed using the \emph{dimer model} --also known as \emph{brane tiling}-- approach. In this framework, a dimer model on a torus is shown to define a relativistic integrable system.

The dimer model provides a graphical way to encode the integrable structure of these systems, where the interactions between particles or variables are represented through a tiling of the plane (or torus) with dimers, which are edges covering vertices in a bipartite graph.

Kenyon's work establishes that any dimer model placed on a torus leads to an integrable system, thereby connecting the combinatorial and geometric aspects of dimer models to the theory of integrable systems.

Let us give a quick review on the dimer model. 

\paragraph{}
A \emph{bipartite graph} $\Gamma=(B,W,E)$ embedded on a oritented 2-manifold $\CalS$ is defined by the following components \cite{kenyon2003introduction}:
\begin{itemize}
    \item {\bf Black nodes} $B$: a finite set of black nodes on $\CalS$, each of which is assigned to the set $B$. 
    \item {\bf White nodes} $W$: a finite set of white nodes on $\CalS$, each of which is assigned to the set $W$. 
    \item {\bf Edges} $E$: A finite set of edges that are embedded as closed intervals on $\CalS$. Each edge connects a black node to a white node. Importantly, edges can only intersect each other at their endpoints.
\end{itemize}
For a bipartite graph $\Gamma=(B,W,E)$ to qualify as a dimer model it must satisfy the following conditions: 
\begin{itemize}
    \item Every equivalent node is on the boundary of $\CalS$.
    \item Every faces of $\Gamma$ is simply-connected. 
\end{itemize}
In this context, we focus on the case where $\CalS = T^2$(the torus). When embedded on the torus, a dimer model is characterized by a \emph{unit cell}. This unit cell has periodicity in both the horizontal and vertical directions, aligning with the periodic boundary conditions of the torus. The boundary of the unit cell represents the two fundamental cycles of the torus:
\begin{itemize}
    \item Horizontal Cycle: Periodic boundary condition in the horizontal direction.
    \item Vertical Cycle: Periodic boundary condition in the vertical direction.
\end{itemize}

\begin{defn}[Perfect matching]
A perfect matching $\BM \subset E$ is a subset of edges such that each node (both black and white) is connected by exactly one edge in $\BM$. The edges in $\BM$ are oriented from black nodes to white nodes by default.
\end{defn}

\begin{defn}[Opposite orientation]
For any perfect matching $\BM$, the notation $-\BM$ refers to the same set of edges but with reversed orientation: edges are now oriented from white nodes to black nodes.
\end{defn}

\begin{defn}[Edge weights assignment]
Each oriented edges $e \in E$ is assigned a weight $w_e$ based on its orientation and its interaction with the boundaries of the unit cell on the torus. The assignment of weights is defined as follows:
\begin{itemize}
    \item Vertical Boundary Crossings:
    \begin{itemize}
        \item Positive Orientation (edge crosses from the left boundary to the right boundary of the unit cell): The weight is $w_e=X$.
        \item Negative Orientation (edge crosses from the right boundary to the left boundary): The weight is $w_e = X^{-1}$.
    \end{itemize}
    \item Horizontal Boundary Crossings:
    \begin{itemize}
        \item Positive Orientation (edge crosses from the bottom boundary to the top boundary of the unit cell): The weight is $w_e = Y$. 
        \item Negative Orientation (edge crosses from the top boundary to the bottom boundary): The weight is $w_e=Y^{-1}$. 
    \end{itemize}
    \item Within the Unit Cell: 
    \begin{itemize}
        \item Edges that do not cross any boundary: The weight is $w_e=1$. 
    \end{itemize}
\end{itemize}
The weight assignment can be organized as
\begin{align}
    w_e = 
    \begin{cases}
        X, & \text{if edge crosses vertical boundary with positive orientation. } \\
        X^{-1}, & \text{if edge crosses vertical boundary with negative orientation. } \\ 
        Y, & \text{if edge crosses horizontal boundary with positive orientation. } \\
        Y^{-1}, & \text{if edgees cross horizontal boundary with negative orientation } \\
        1, & \text{The edge lies within the unit cell. }
    \end{cases}
\end{align}
\end{defn}

The weight of a perfect matching $\BM$ is a product over the weights of all its edges:  
\begin{align}
    W[\BM] = \prod_{e\in \BM} w_e
\end{align}
The spectral curve $\CalC$ of the dimer graph $\Gamma$ is the ensemble of all possible perfect matchings' weight: 
\begin{align}
    \CalC = \left\{ (X,Y) \in \BC^2 \left| W(X,Y) = \sum_{\{\BM\}} W[\BM] = 0 \right. \right\}
\end{align}

The dual graph of the dimer graph is a planar, periodic quiver \cite{Franco:2005rj}. Quiver gauge theories described by dimer graphs arise from the worldvolume of a stack of D3-branes probing a single Calabi-Yau (CY) 3-fold. The connection between dimer models and quivers has trivialized the determination of the Calabi-Yau geometry \cite{Franco:2005rj,Benvenuti:2004wx,Huang:2020neq}. 
This curve happens to be the mirror curve of the Calabi-Yau 3-fold. 
It can also be obtained by considering the \emph{Kasteleyn matrix} of the dimer \cite{treumann2019kasteleyn}. The Kasteleyn matrix is a weighted adjacency matrix of the graph.
A toric diagram $\ET$ can be associated to the dimer graph $\Gamma$ based on its curve $W(X,Y)$ \cite{Hanany:2005ve}.

We need some additional definition to establich relation between dimer graph and integrable models \cite{Kenyon2011dimers}. 
\begin{defn}[Reference perfect matching]
In this note we always choose the perfect matching that corresponds to the furthest point in the positive $X$ axis on the corresponding toric diagram $\ET$ as the prefect matching $\BM_{\rm ref}$. 
\end{defn}

\begin{defn}[1-loop]
A 1-loop is a path connecting from a node in the graph to the same node which passes through the boundary of the unit cell. 
\end{defn}
To systematically construct 1-loop, one can consider the difference between a general perfect matching and the reference perfect matching $\BM - \BM_{\rm ref} $. 

\begin{defn}[$n$-loop]
    The $n$-loop is product of $n$ non-overlapping 1-loops. 
\end{defn}

\begin{figure}
    \centering
    \includegraphics[width=0.75\textwidth]{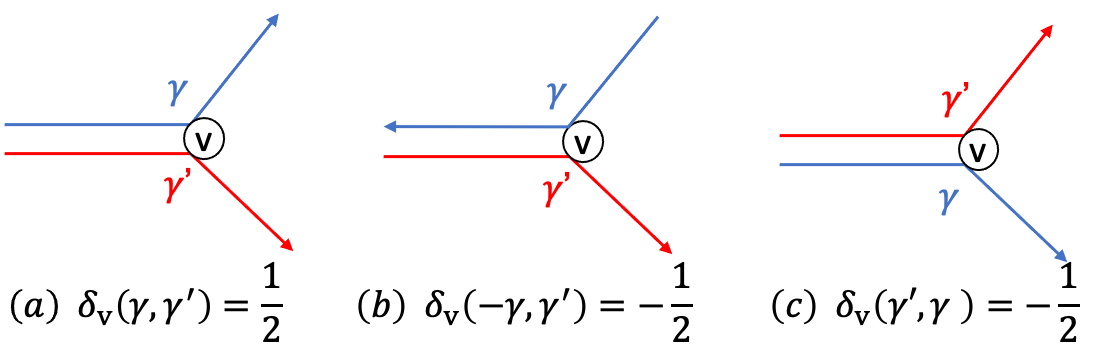}
    \caption{An illustration of $\d_v(\g,\g')$. }
    \label{fig:delta-vertex}
\end{figure}

\begin{defn}[Poisson commutation]\label{def:commu}
    Let $\g$ and $\g'$ be two oriented loops on the dimer graph $\Gamma$. The Poisson commutation relation between the loops are defined as
\begin{align}
    \{\g,\g'\} = \epsilon_{\g\g'} \g \g'
\end{align}
where
\begin{align}
   \epsilon_{\g\g'}  = \sum_{v} {\rm sgn}(v) \d_v(\g,\g').
\end{align}
\end{defn}
Here $\sgn(v)=1$ for the black node and $\sgn(v)=-1$ for the white node. $\d_v$ is a skewsymmetric bilinear form satisfying 
\[
    \d_v(\g,\g') = -\d_v(\g',\g) = -\d_v(-\g,\g') \in \frac{1}{2} \BZ
\]
as illustrated in Fig.~\ref{fig:delta-vertex}. 

\begin{defn}[Hamiltonian]\label{def:Hamil}
    The $n$-th Hamiltonian of the dimer graph is defined as the sum over all the $n$-loops. 
\end{defn}

\begin{thm}
    The poisson commutation \ref{def:commu} defines a classical integrable systems, where the conserving integrable of motions are the Hamiltonians \ref{def:Hamil} \cite{kenyon2003introduction,Hanany:2005ve}.
\end{thm}

\paragraph{}

\begin{figure}
    \centering
    \includegraphics[width=0.5\textwidth]{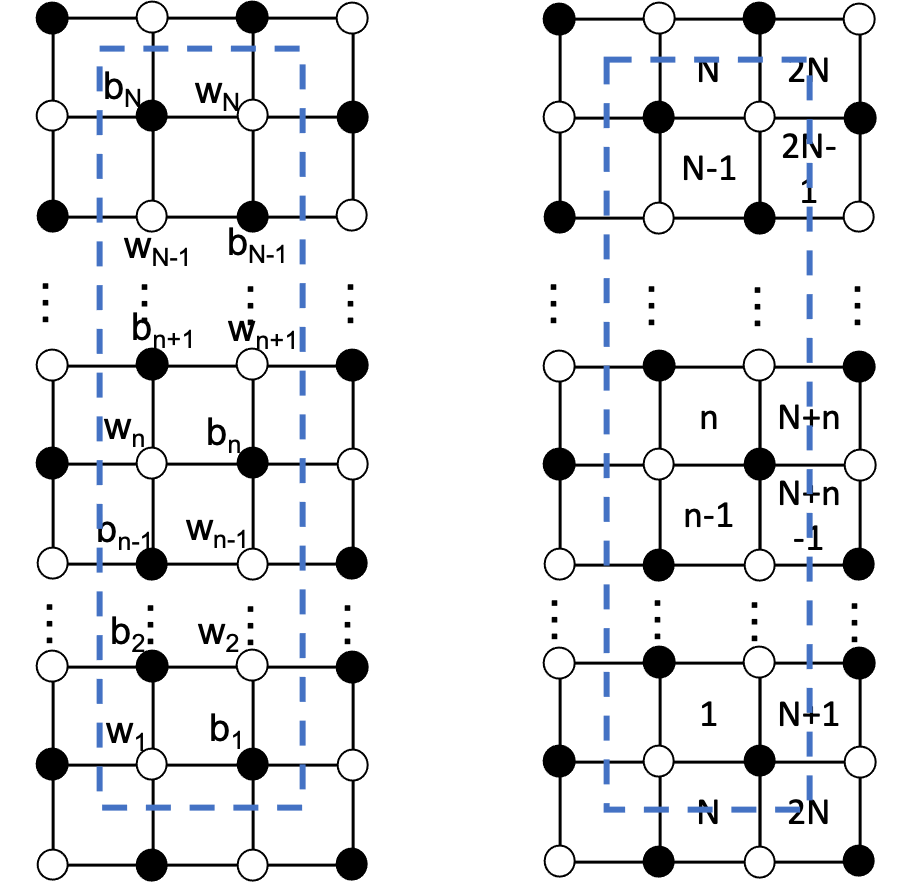}
    \caption{The brane tiling for $Y^{N,0}$ dimer model with $N$ even. A unit cell is encircled by the dashed blue line. This dimer is associated to the $\hat{A}_{N-1}$ relativistic Toda lattice.}
    \label{fig:YN0}
\end{figure}

The dimer model associated to the type A RTL is known as $Y^{N,0}$ model. It is constructed by periodic square diagrams. The shape of the unit cell depends on whether $N$ is even (rectangular) or odd (rhombus). For the purpose of this paper, here we consider the case $N$ being even. Each unit cell consists two columns of $N$ squares along with $N$ white and black vertexes $\tw_n$, $\tb_n$, $n=1,\dots,N$. See Figure.~\ref{fig:YN0} for illustration.

The first Hamiltonian of the $Y^{N,0}$ dimer coincide with the $\hat{A}_{N-1}$ RTL Hamiltonian once the 1-loops are expressed in terms of the canonical coordinates. 
The Lax matrix of $\hat{A}_{N-1}$ RTL can be constructed from the Kasteleyn matrix. 
We will not get in to the detail here but refer the interested reader to Appendix.~\ref{sec:type A dimer} and \cite{Lee:2023wbf,Eager:2011dp} for more detail. 

\subsection{Introduce impurities}

The relation between the toric diagram and dimer graph works in both directions. A dimer graph $\Gamma=(B,W,E)$ can be constructed based on a given toric diagram $\ET$ \cite{Eager:2011dp,Hanany:2005ve}. 
In some cases multiple dimer graphs can be constructed based on a single toric diagram \cite{Lee:2023wbf}. 

We use the following two facts to construct the dimer model for type D RTL:
\begin{itemize}
    \item Seiberg-Witten curve of supersymmetric Yang-Mills theory of gauge group $G$ coincides with spectral curve of RTL of Lie algebra $\fg = \text{Lie}(G)$. 
    \item $SO(2N)$ gauge theory shares the same toric diagram with $SU(2N)$ gauge theory with 8 fundamental matters fine-tuned and can be embedded into $SU(2N)$ theory with folding to reduce the degree of freedom. 
\end{itemize}

The dimer graph associated to the toric diagram for $SU(2N)+8F$ can be constructed by introducing impurity in the $Y^{2N-4,0}$ model (corresponding to the $A_{2N-4}$ RTL). This is indicated by the slopes of the four long edges of the toric diagram. The $D_N$ dimer can then be obtained through folding, which eliminates half of the degrees of freedom.

\begin{figure}
    \centering
    \includegraphics[width=0.5\textwidth]{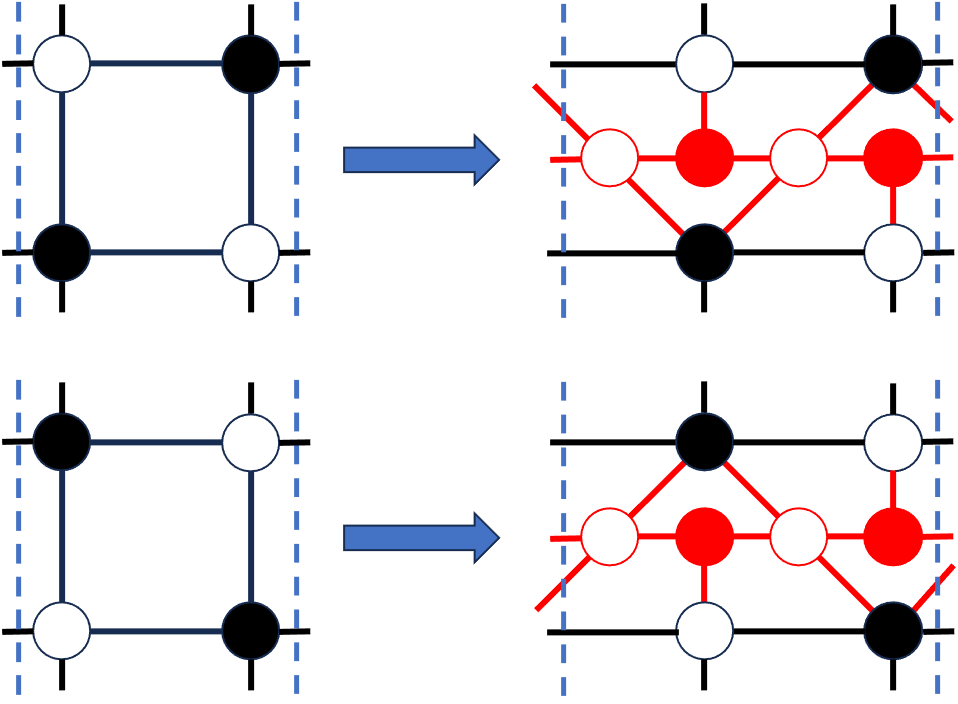}
    \caption{The vertical lines in the toric diagram $\ET$ introduces impurity to the square dimer graph. The newly introduced red nodes modifies the edges inside the square. The boundary of the unit cell is indicated by blue dashed lines.}
    \label{fig:D-impurity}
\end{figure}

The 8 short vertical lines in the toric diagram $\ET$ introduce impurity to the square dimer in the following way: When $N\geq 4$, we pick $4$ among $2N-4$ squares in the unit cell the $n_i$-th square, $i=1,2,3,4$,  $n_i \in \{1,\dots,2N-4\}$, to introduce impurity. 
Two pairs of black and white nodes, which we will denote by $\tilde\tw_{n_i}^{(1,2)}$ and $\tilde\tb_{n_i}^{(1,2)}$ are added in the middle of the $n_i$-th square in the order white-black-white-black starting from the left of the unit cell. 
The vertical edges of $n_i$-th squares are removed and replaced by 
\begin{itemize}
    \item Horizontal lines connecting the newly introduced black and white nodes, which also extended to outside of the the unit cell. 
    \item Two vertical lines connecting the newly introduced black nodes $\tb_{n_i}^{(l)}$, $l=1,2$ to the square's white nodes $\tw_{n_i-\frac{1}{2} + \frac{(-1)^{n_i}}{2} }$.  
    \item Two slide edges connecting $\tw_{n_i}^{l}$, $l=1,2$, to $\tb_{n_i}$ and $\tb_{n_i-1}$. 
\end{itemize}
See Fig.~\ref{fig:D-impurity} for illustration. 

\begin{figure}
    \centering
    \includegraphics[width=0.5\textwidth]{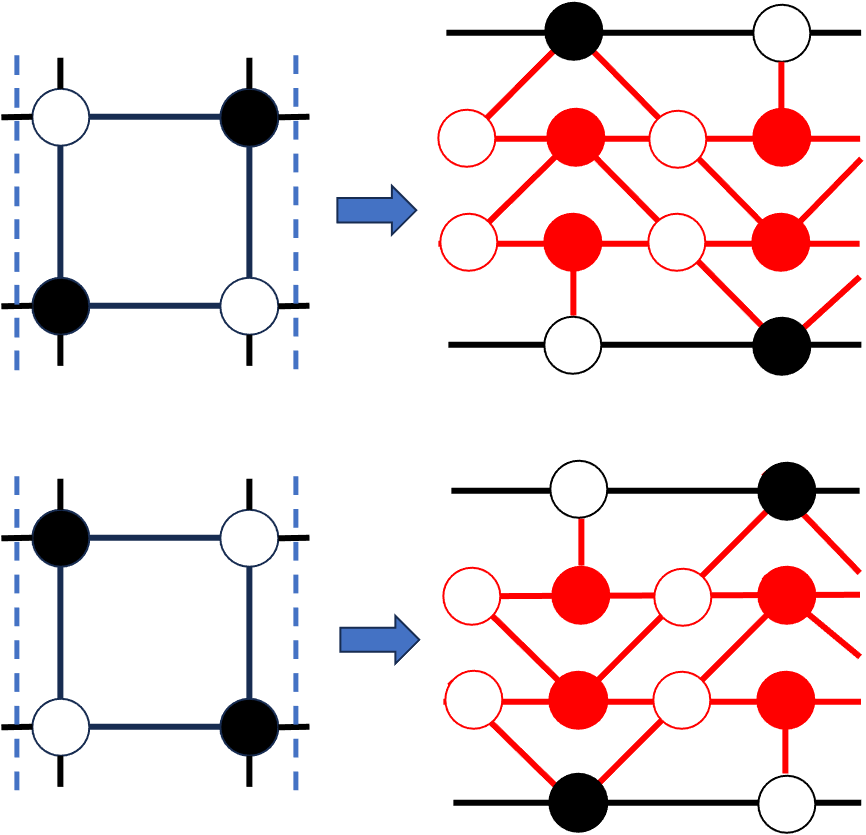}
    \caption{The vertical lines in the toric diagram $\ET$ introduces double impurity to the square dimer graph. The newly introduced red nodes modifies the edges inside the square. The boundary of the unit cell is indicated by blue dashed lines.}
    \label{fig:D-impurity-2}
\end{figure}

Alternatively, one can place double impurity to a single square in the $A_{2N-4}$ dimer. Four pairs of white and black nodes are added to the middle of the square. See Fig.~\ref{fig:D-impurity-2} for illustration. 

The final dimer graph will have exactly $2N-4+4\times 2 = 2N+4$ pairs of black and white nodes distributed on exactly $2N$ rows. 

\subsection{$D_4$}\label{sec:D4}

\begin{figure}
    \centering
    \includegraphics[width=0.6\textwidth]{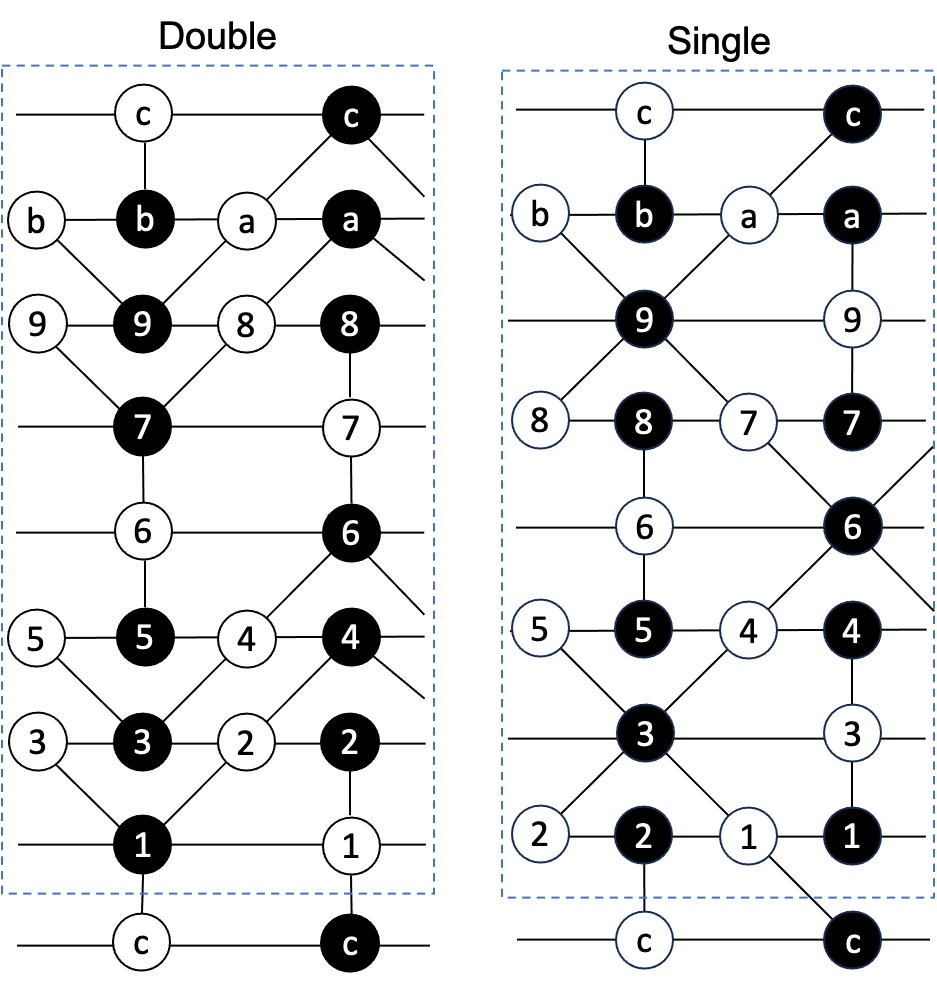}
    \caption{Two bipartite graphs  generated by toric diagram for $D_4/A_7+8F$ in Fig.~\ref{fig:so(8) - web}. The left graph is constructed by placing double impurities at the first and third square in the $A_4$ dimer graph. The right one place single impurities in each of the square in the $A_4$ dimer graph. }
    \label{fig:SU(8)+8F-dimer}
\end{figure}

The dimer graph for $D_4$ system can be constructed by introducing impurities into the $A_4$ dimer graph. The impurity can be introduced in the form of 4 single impurities or as 2 double impurities. Figure~\ref{fig:SU(8)+8F-dimer} illustrates these two different bipartite graphs. More general cases can also be considered, which we will discuss in future work.
 
The Kastelyne matrix $K_{4,D}$ of the double impurity bipartite graph in Fig.~\ref{fig:SU(8)+8F-dimer} is:

\begin{tabular}{c|c c c c c c c c c c c c}
     & 1 & 2 & 3 & 4 & 5 & 6 & 7 & 8 & 9 & a & b & c \\
     \hline
    1 & $h_{1r}-\frac{h_{1l}}{X}$ & $s_{12}$ & $s_{13}$ & 0 & 0 & 0 & 0 & 0 & 0 & 0 & 0 & $\frac{v_{1c}}{Y}$ \\
    2 & $v_{21}$ & $-h_2$ & $h_{23}X$ & 0 & 0 & 0 & 0 & 0 & 0 & 0 & 0 & 0 \\
    3 & 0 & $h_{32}$ & $-h_3$ & $s_{34}$ & $s_{35}$ & 0 & 0 & 0 & 0 & 0 & 0 & 0 \\
    4 & 0 & $s_{42}$ & $s_{43}X$ & $-h_4$ & $h_{45}X$ & 0 & 0 & 0 & 0 & 0 & 0 & 0 \\
    5 & 0 & 0 & 0 & $h_{54}$ & $-h_5$ & $v_{56}$ & 0 & 0 & 0 & 0 & 0 & 0 \\
    6 & 0 & 0 & 0 & $s_{64}$ & $s_{65}X$ & $h_{6r}X - {h_{6l}}$ & $v_{67}$ & 0 & 0 & 0 & 0 & 0 \\
    7 & 0 & 0 & 0 & 0 & 0 & $v_{76}$ & $h_{7r}-\frac{h_{7l}}{X}$ & $s_{78}$ & $s_{79}$ & 0 & 0 & 0 \\
    8 & 0 & 0 & 0 & 0 & 0 & 0 & $v_{87}$ & $-h_8$ & $h_{89}X$ & 0 & 0 & 0 \\
    9 & 0 & 0 & 0 & 0 & 0 & 0 & 0 & $h_{98}$ & $-h_9$ & $s_{9a}$ & $h_{9b}$ & 0 \\
    a & 0 & 0 & 0 & 0 & 0 & 0 & 0 & $s_{a8}$ & $s_{a9}X$ & $-h_a$ & $h_{ab}X$ & 0 \\
    b & 0 & 0 & 0 & 0 & 0 & 0 & 0 & 0 & 0 & $h_{ba}$ & $-h_b$ & $v_{bc}$ \\
    c & $v_{c1}Y$ & 0 & 0 & 0 & 0 & 0 & 0 & 0 & 0 & $s_{ca}$ & $s_{cb}X$ & $h_{cr} X- {h_{cl}}$ 
\end{tabular}
In particular, when setting all parameter (beside $X$ and $Y$) to 1 gives
\begin{align}\label{eq:det-K4D-all1}
\begin{split}
    \frac{\det K_{4,D}}{X^2} 
    & = X^4 + \frac{1}{X^4} - 36 \left(X^3 + \frac{1}{X^3}\right) + 406 \left(X^2 + \frac{1}{X^2}\right) - 1556 \left(X + \frac{1}{X}\right) + 2402 \\
    & \quad - \left( Y + \frac{1}{Y} \right) \left( X^2 - 4X + 6 - \frac{4}{X} + \frac{1}{X^2} \right)
\end{split}
\end{align}

\begin{figure}
    \centering
    \includegraphics[width=1\textwidth]{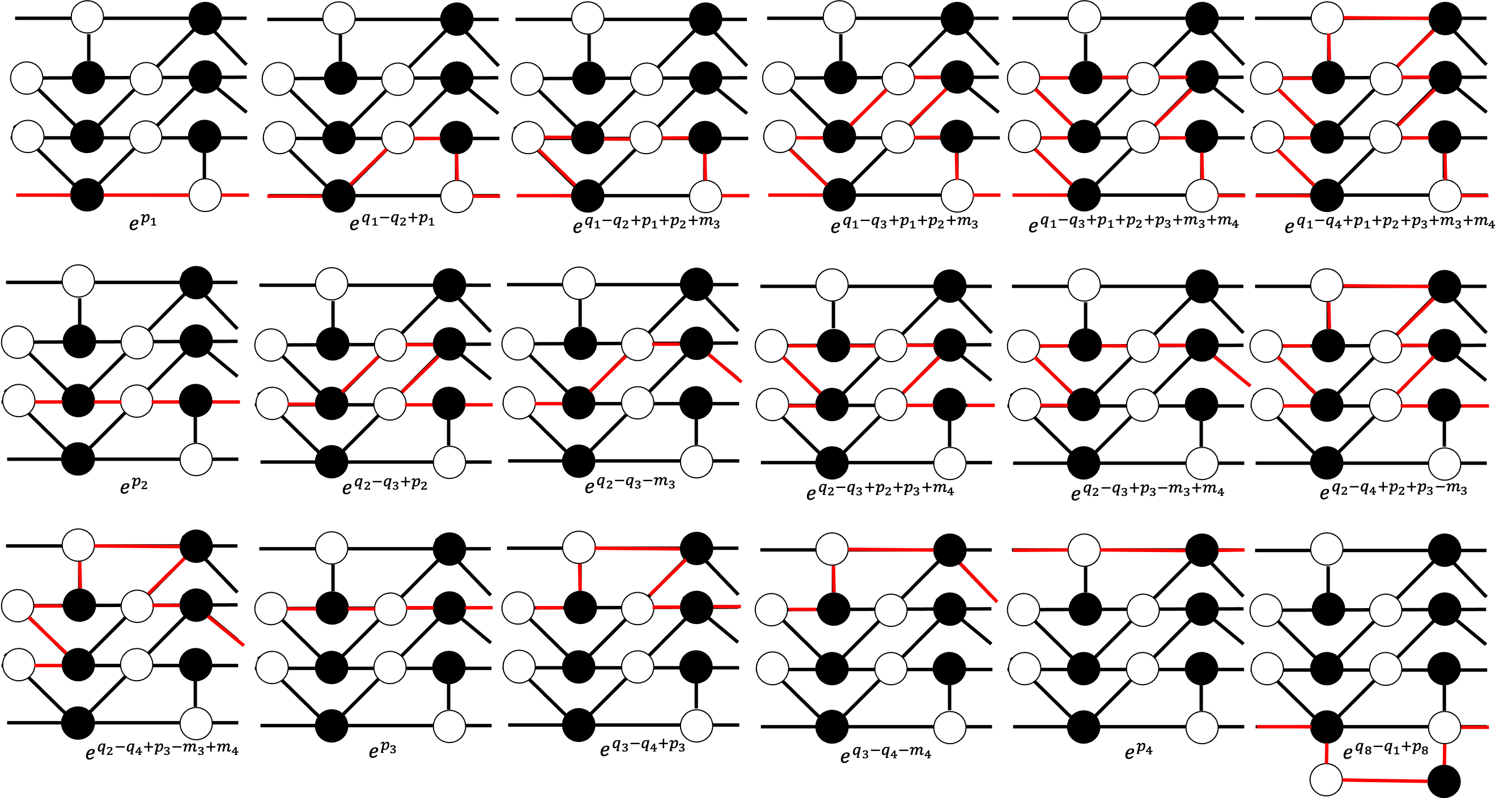}
    \caption{18 1-loops in the double impurity $SU(8)+8F$ dimer graph.}
    \label{fig:SU(8)+8F-1-loop}
\end{figure}

\begin{figure}
    \centering
    \includegraphics[width=0.6\textwidth]{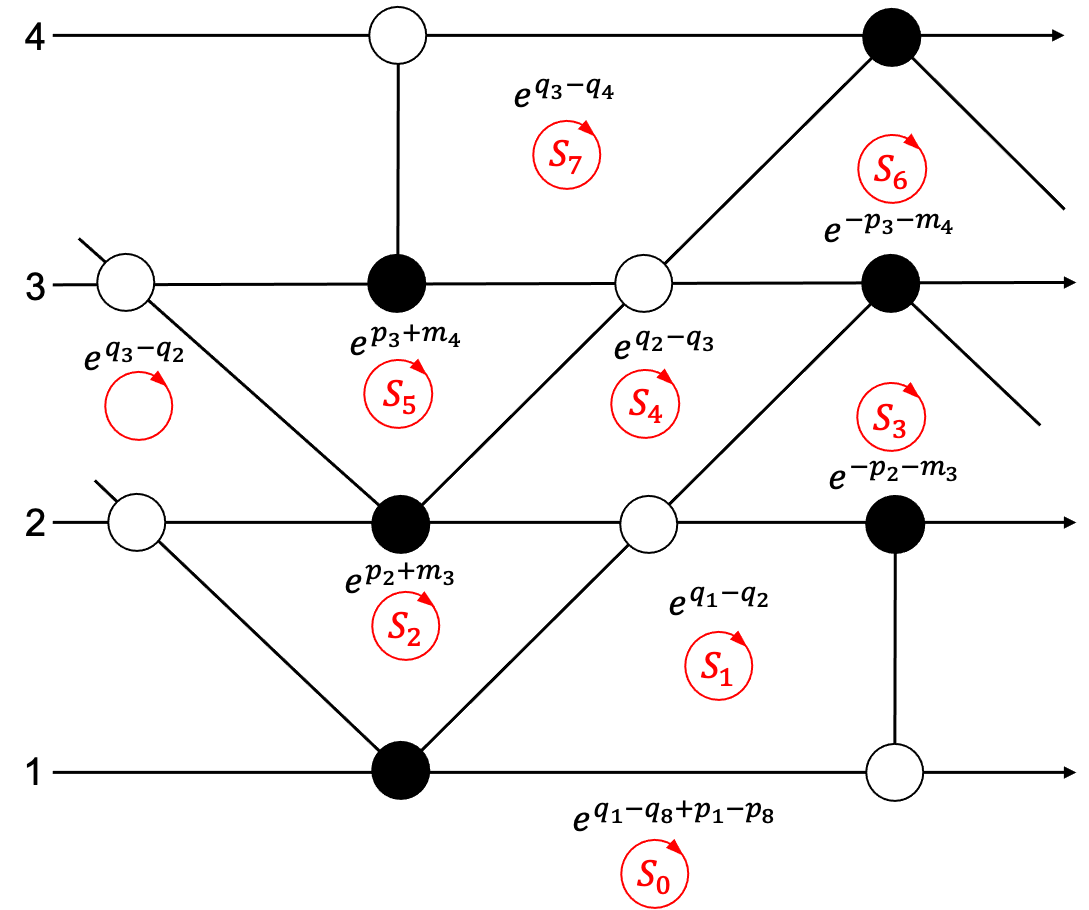}
    \caption{Basic square loops in the modified unit cell in double impurity dimer. The four masses are encoded in the dimer graph through the square loops and impurity straight lines. }
    \label{fig:SU(8)+8F-1-loop-fund}
\end{figure}

We choose the reference perfect matching such that each dimer covers $\tw_i\to\tb_i$ with all arrows going rightward. The number of 1-loops is the coefficient of $-X^3$ in \eqref{eq:det-K4D-all1}: 36. These 1-loops are distributed across the 8 horizontal lines:
\begin{itemize}
    \item Lines with $\tw_1$ and $\tw_7$: 7 in total; 
    \item Lines with $\tw_3$ and $\tw_9$: 7 in total;
    \item Lines with $\tw_5$ and $\tw_{b}$: 3 in total; 
    \item Lines with $\tw_6$ and $\tw_{c}$: 1 in total;
\end{itemize}
See Fig.~\ref{fig:SU(8)+8F-1-loop} for illustration of the 1-loops. Most of the 1-loops can be built from straight lines $d_j$, $j=1,2,3,4$, and square loops $S_n$, $n=1,\dots,7$. See Fig.~\ref{fig:SU(8)+8F-1-loop-fund} for an illustration. 

The double impurity dimer is constructed from the ${A}_7+8F$ toric diagram Fig.\ref{fig:so(8) - web}. 
The masses $m_f$, $f=1,\dots,8$ are fine-tuned to 
\begin{align}
    m_1=m_2 = 0, \ m_3 = -m_4 = -\ri \pi, \ m_5=m_6 = 0, \ m_7 = -m_8 = -\ri \pi. 
\end{align}

The first Hamiltonian for the double impurities -- the dimer graph on the left of Fig.~\ref{fig:SU(8)+8F-dimer} --- is the sum over all 1-loops:
\begin{align}
\begin{split}
    H_1 = & \ e^{\tp_1} \left[ 1 + e^{\tq_1-\tq_2} - e^{\tq_1-\tq_2+\tp_2} - e^{\tq_1-\tq_3+\tp_2} + e^{\tq_1-\tq_3+\tp_2+\tp_3} + e^{\tq_1-\tq_4+\tp_2+\tp_3} \right] \\
    & + e^{\tp_2} \left[ 1 + e^{\tq_2-\tq_3} - e^{\tq_2-\tq_3-\tp_2} - e^{\tq_2-\tq_3+\tp_3} + e^{\tq_2-\tq_3 - \tp_2+\tp_3} - e^{\tq_2-\tq_4 +\tp_3} + e^{\tq_2-\tq_4-\tp_2+\tp_3} \right] \\
    & + e^{\tp_3} \left[ 1 + e^{\tq_3-\tq_4} - e^{\tq_3-\tq_4-\tp_3} \right] \\
    & + e^{\tp_4} \left[ 1 + e^{\tq_4-\tq_5} \right] \\
    & + e^{\tp_5} \left[ 1 + e^{\tq_5-\tq_6} - e^{\tq_5-\tq_6+\tp_6} - e^{\tq_5-\tq_7+\tp_6} + e^{\tq_5-\tq_7+\tp_6+\tp_7} + e^{\tq_5-\tq_8+\tp_6+\tp_7} \right] \\
    & + e^{\tp_6} \left[ 1+ e^{\tq_6-\tq_7} - e^{\tq_6-\tq_7-\tp_6} - e^{\tq_6-\tq_7+\tp_7} + e^{\tq_6-\tq_7 - \tp_6+\tp_7} - e^{\tq_6-\tq_8 +\tp_7} + e^{\tq_6-\tq_8-\tp_6+\tp_7} \right] \\
    & + e^{\tp_7} \left[1 + e^{\tq_7-\tq_8} - e^{\tq_7-\tq_8-\tp_7} \right] \\
    & + e^{\tp_8}  \left[ 1+ e^{\tq_8-\tq_1} \right].  \\
\end{split}
\end{align}
The seventh Hamiltonian is the sum over all 7-loops: 
\begin{align}
\begin{split}
    e^{-\tp_\text{tot} }H_7 = 
    & \ e^{-\tp_1} \left[ 1 + e^{\tq_8-\tq_1}\right] \\
    & + e^{-\tp_2} \left[ 1 + e^{\tq_1-\tq_2} - e^{\tq_1-\tq_2+\tp_2} \right] \\
    & + e^{-\tp_3} \left[ 1 + e^{\tq_2-\tq_3} - e^{\tq_2-\tq_3-\tp_2} - e^{\tq_2-\tq_3+\tp_3} + e^{\tq_2-\tq_3-\tp_2+\tp_3} - e^{\tq_1-\tq_3-\tp_2} + e^{\tq_1-\tq_3-\tp_2+\tp_3} \right] \\
    & + e^{-\tp_4} \left[ 1+ e^{\tq_3-\tq_4} - e^{\tq_3-\tq_4-\tp_3} - e^{\tq_2-\tq_4-\tp_3} + e^{\tq_2-\tq_4-\tp_3-\tp_2} + e^{\tq_1-\tq_4-\tp_3-\tp_2} \right] \\
    & + e^{-\tp_5} \left[ 1 + e^{\tq_4-\tq_5} \right] \\
    & + e^{-\tp_6} \left[ 1 + e^{\tq_5-\tq_6} - e^{\tq_5-\tq_6+\tp_6} \right] \\
    & + e^{-\tp_7} \left[ 1 + e^{\tq_6-\tq_7} - e^{\tq_6-\tq_7-\tp_6} - e^{\tq_6-\tq_7+\tp_7} + e^{\tq_6-\tq_7-\tp_6+\tp_7} - e^{\tq_5-\tq_7-\tp_6} + e^{\tq_5-\tq_7-\tp_6+\tp_7} \right] \\
    & + e^{-\tp_8} \left[ 1+ e^{\tq_7-\tq_8} - e^{\tq_7-\tq_8-\tp_7} - e^{\tq_6-\tq_8-\tp_7} + e^{\tq_6-\tq_8-\tp_7-\tp_6} + e^{\tq_5-\tq_8-\tp_7-\tp_6} \right] \\
\end{split}
\end{align}

To fold from $SU(8)+8F$ to $SO(8)$, we need to eliminate half of the degree of freedom to make $H_1=H_7$.  
This can be done by setting
\begin{align}\label{eq:folding-solution}
\begin{split}
    & \tq_1=-\tq_4, \ \tp_1 = -\tp_4, \ \tq_2 = -\tq_3, \ \tp_2=-\tp_3, \\
    & \tq_5=-\tq_8, \ \tp_5 = -\tq_8, \ \tq_6 = -\tq_7, \ \tp_6=-\tp_7. 
\end{split}
\end{align}
The first and seventh Hamiltonian become
\begin{align}
\begin{split}
    H_1 = H_7
    = & \ 2\cosh \tp_1 + 2\cosh\tp_2 + 2\cosh\tp_7 + 2\cosh \tp_8 \\
    & + e^{\tq_1-\tq_2} (e^{\tp_1}+e^{-\tp_2} - 1 - e^{\tp_1+\tp_2}) \\
    & + e^{\tq_7-\tq_8} (e^{\tp_7}+e^{-\tp_8} - 1 - e^{-\tp_7-\tp_8}) \\
    & + e^{\tq_8-\tq_1} ( e^{-\tp_1} + e^{\tp_8}) \\
    & + e^{\tq_1+\tq_2} (e^{\tp_1} + e^{-\tp_2} - 1-e^{\tp_1+\tp_2}) \\
    & + e^{-\tq_7-\tq_8} (e^{\tp_7}+e^{-\tp_8} - 1 - e^{-\tp_7-\tp_8}) \\
    & + e^{2\tq_1+\tp_1} + e^{2\tq_2} \left[ e^{\tp_2} - 2 + e^{-\tp_2} \right] + e^{-2\tp_7} \left[ e^{\tp_7} - 2 + e^{-\tp_7} \right] + e^{-2\tq_8-\tp_8} \\
\end{split}
\end{align}
We would like to point out that the identification \eqref{eq:folding-solution} causes the Poisson commutation relation defined for shared edges to potentially fail when there are square loops between the double impurity lines ($S_3$, $S_4$, and $S_5$ in Fig.~\ref{fig:SU(8)+8F-1-loop-fund}). After imposing the solution, the Poisson commutation relation involving $S_4$ are doubled despite sharing only one edge: 
\[
    \frac{\{S_4,S_2\}}{S_4S_2} = \frac{\{S_5,S_4\}}{S_5S_4} = \frac{\{S_3,S_4\}}{S_3S_4} = \frac{\{S_4,S_6\}}{S_4S_6} = \frac{\{S_4,e^{\tp_2}\}}{S_4e^{\tp_2}} = 2. 
\]
To fully match $\hat{D}_4$ Hamiltonian, we consider the following change of variables: 
$\tq_1 \to \tq_1 - \frac{\tp_1}{2}$, $\tq_8 \to \tq_8 - \frac{\tp_8}{2}$ and 
\begin{align}\label{eq:change of variable}
\begin{split}
    & e^{\tq_2} \to  \frac{\cosh \frac{\tp_2}{2} }{\sinh \tq_2}, \quad e^{\tp_2} \to \frac{\cosh\frac{\tp_2-2\tq_2}{2}}{\cosh \frac{\tp_2+2\tq_2}{2}} ; 
    \quad e^{\tq_7} \to \frac{\sinh\tq_7}{\cosh\frac{\tp_7}{2}}, \quad e^{\tp_7} \to \frac{\cosh\frac{\tp_7+2\tq_7}{2}}{\cosh\frac{\tp_7-2\tq_7}{2}}
\end{split}
\end{align}
The change of variables are canonical, i.e. preserve the Poisson commutation relation. This can be checked by direct computation. See Appendix.~\ref{sec:change-of-variables} for computational detail. 
Finally we modify $S_{1,7}$ by
\begin{align}\label{eq:mod-square}
    S_1 \to S_1 \frac{\cosh\frac{\tp_2}{2}}{\sinh\frac{\tp_2}{2}}, \quad S_7 \to S_7 \frac{\sinh\frac{\tp_2}{2}}{\cosh\frac{\tp_2}{2}}. 
\end{align}
The same modification is also applied to the other double impurity. 
The first Hamilton $H_1$ after the change of canonical coordinate becomes:

\begin{align}\label{eq:H-D4-dimer}
\begin{split}
    H_1 = & \ 2\cosh \tp_1 + 2\cosh \tp_2 + 2\cosh\tp_7 + 2\cosh\tp_8 \\
    & + e^{\tq_1-\tq_2} 2\cosh \frac{\tp_1+\tp_2}{2} + e^{\tq_8-\tq_1} 2\cosh \frac{\tp_8+\tp_1}{2} + e^{\tq_7-\tq_8} 2\cosh \frac{\tp_7+\tp_8}{2} \\
    & + e^{\tq_1+\tq_2} 2\cosh \frac{\tp_1-\tp_2}{2} + e^{-\tq_7-\tq_8} 2\cosh \frac{\tp_7-\tp_8}{2} + e^{2\tq_1} + e^{-2\tq_8} \\
\end{split}
\end{align}
After renaming $1\to 3$, $2\to4$, $8\to2$, and $7\to1$, we recover \eqref{def:H-RTL-D} with coupling $g=1$
\[
    H_1 = {\rm H}_{D_4}.
\]

\begin{figure}
    \centering
    \includegraphics[width=0.9\textwidth]{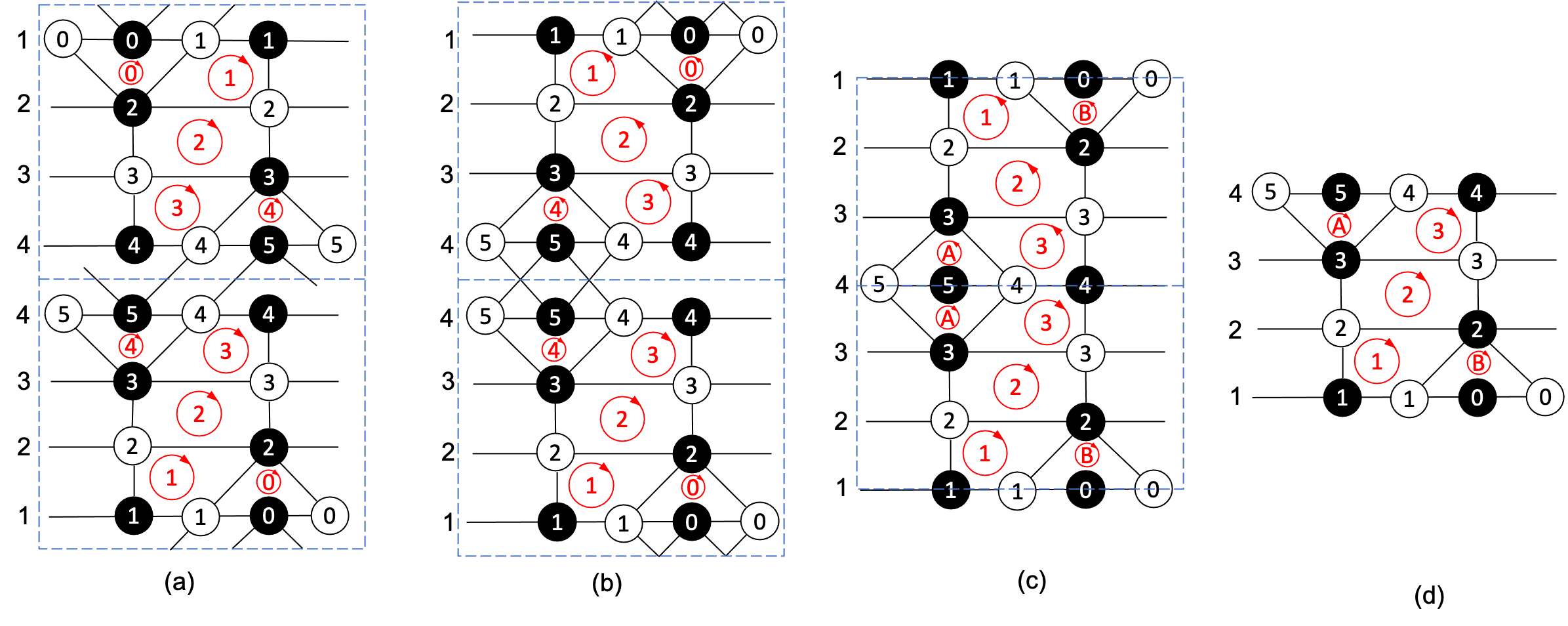}
    \caption{Identifying the vertexes and faces in the double impurity dimer Fig.~\ref{fig:SU(8)+8F-1-loop-fund} based on Eq.~\eqref{eq:folding-solution} in figure (a).
    In (b) the top half unit cell is reflected horizontally. 
    The change of canonical coordinate \eqref{eq:change of variable} modifies the two 1st and two 4th line. This further allows folding of the top half unit cell to obtain (d). 
      }
    \label{fig:SU(8)-D4 folding}
\end{figure}

On the double impurity dimer graph, we identify the vertices and faces in the double impurity $SU(8)+8F$ dimer graph according to the solution \eqref{eq:folding-solution}. The unit cell described by \eqref{eq:folding-solution}can be composed of two identical half unit cells, with one rotated by 180 degrees and placed on top of the other (the identification is based on the level of the square loop). 
This identification across the two half unit cells allows for folding. The edges connecting a node $A$ in one half unit cell to a node $B$ in the other half unit cell are removed according to the change of variable \eqref{eq:change of variable}. See Fig.~\ref{fig:SU(8)-D4 folding} for an illustration. 
This modification further affects the Poisson commutation relation between the 1-loops sharing the double edges.

We obtain the dimer graph shown in Fig.~\ref{fig:D4 dimer}. 
To construct the 1-loops, we choose the perfect matching on the $D_4$ dimer to be horizontal edges with a white node to the left of the black node. There are 9 1-loops based on this reference matching.
Next, we rotate the dimer graph with the 1-loops by 180 degrees. This rotation results in another 9 1-loops on the folded patch. Note that rotating by 180 degrees is equivalent to using a different reference perfect matching, where the white nodes are on the right of the black nodes on each horizontal edge.

\begin{figure}
    \centering
    \includegraphics[width=1\textwidth]{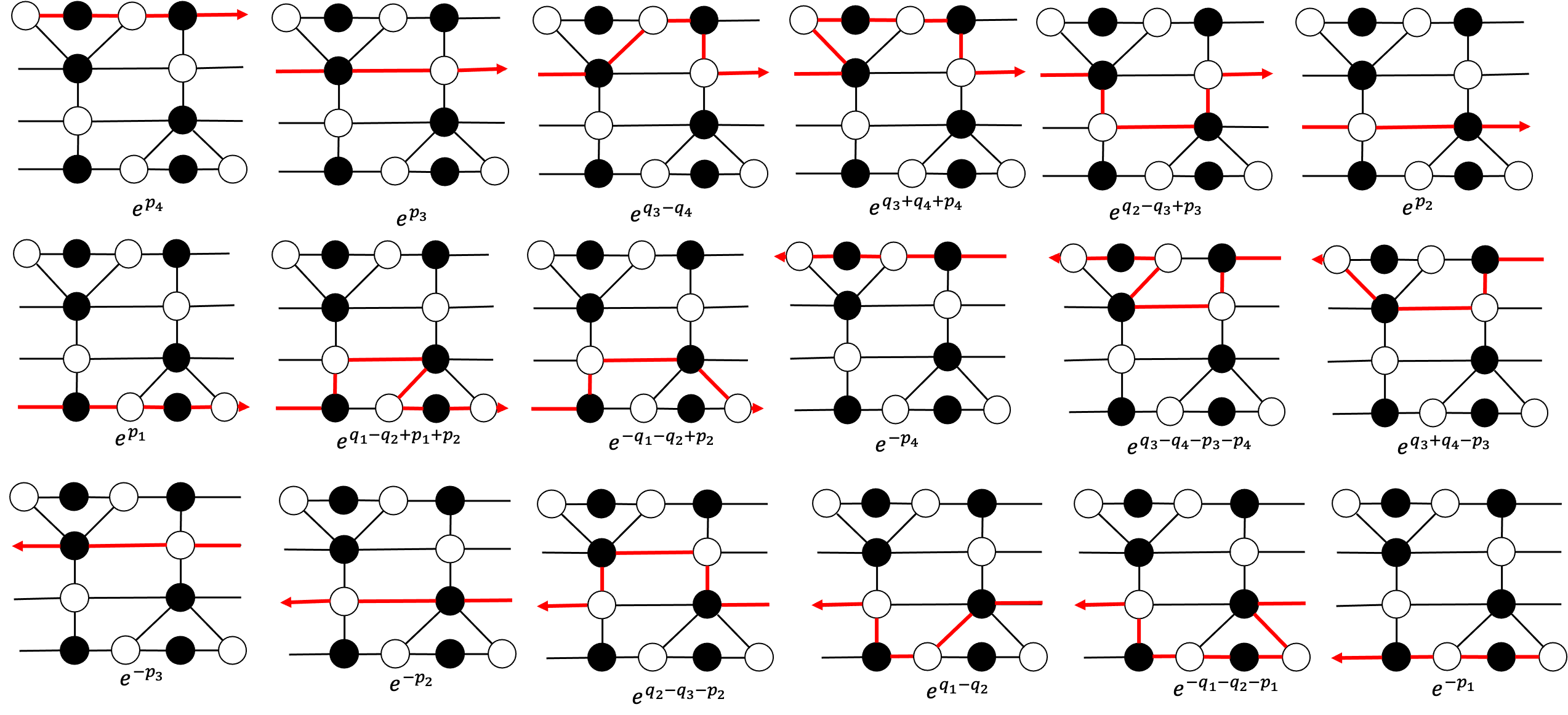}
    \caption{There are 18 1-loops in the $D_4$ dimer graph. The 6 1-loops on the top line and the 3 on the left in the middle are obtained from the reference perfect matching. The remaining 9 1-loops (3 on the right in the middle and 6 on the bottom) are obtained by rotating the dimer by 180 degrees or using an alternative perfect matching. Additionally, there are two more 1-loops that come from connections between the two half unit cells. }
    \label{fig:D4 dimer 1-loop}
\end{figure}

We assign the face loops in the $D_4$ dimer to obey the Poisson commutation relation based on the shared edges and horizontal 1-loops in Fig.~\ref{fig:D4 dimer}. This is sufficient to determine all $9+9=18$ 1-loops in Fig.~\ref{fig:D4 dimer 1-loop}:

\begin{figure}
    \centering
    \includegraphics[width=0.5\textwidth]{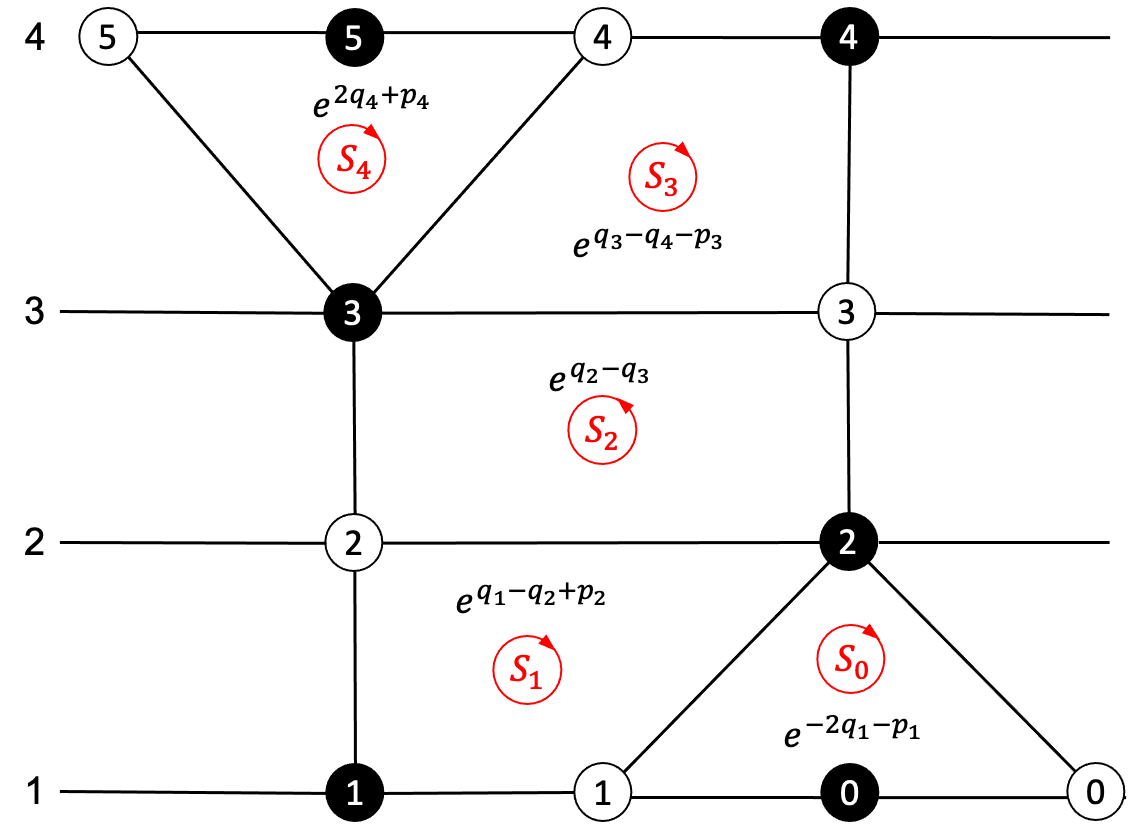}
    \caption{Half of dimer graph for $D_4$ theory, obtained from folding $SU(8)+8F$ with double impurity Fig.~\ref{fig:SU(8)+8F-dimer}. The full dimer is two copies of the half-dimer, with the second rotated by 180 degree.}
    \label{fig:D4 dimer}
\end{figure}

The first Hamiltonian of the dimer graph Fig.~\ref{fig:D4 dimer} is the sum over all 1-loops plus the two additional terms arising from connecting the two patches (top right of Fig.~\ref{fig:SU(8)+8F-1-loop}): 
\begin{align}
\begin{split}
    H_1 = & \ 2 \cosh \tp_1 + 2 \cosh \tp_2 + 2 \cosh \tp_3 + 2\cosh \tp_4 \\
    & + e^{\tq_1-\tq_2} (1 + e^{\tp_1+\tp_2}) + e^{\tq_2-\tq_3} (e^{\tp_3} + e^{-\tp_2}) + e^{\tp_3-\tp_4} (e^{\tp_4}+e^{\tp_3}) \\
    & + e^{-\tq_1-\tq_2} (e^{-\tp_1}+e^{\tp_2}) + e^{\tq_3+\tq_4} (e^{\tp_4}+e^{\tp_3}) + e^{-2\tq_2} + e^{2\tq_3}
\end{split}
\end{align}
Change the canonical coordinate by
\[
    \tq_{2,3}= \tq_{2,3} + \frac{\tp_{2,3}}{2}, \quad \tq_{1,4} = \tq_{1,4} - \frac{\tp_{1,4}}{2}
\]
and we identify the first Hamiltonian as the $D_4$ relativistic Toda lattice ${\rm H}_{D_4}$ in \eqref{def:H-RTL-D}:
\[
    H_1 = {\rm H}_{D_4}. 
\]

\subsection{$D_N$}\label{sec:DN}

\begin{figure}
    \centering
    \includegraphics[width=0.3\textwidth]{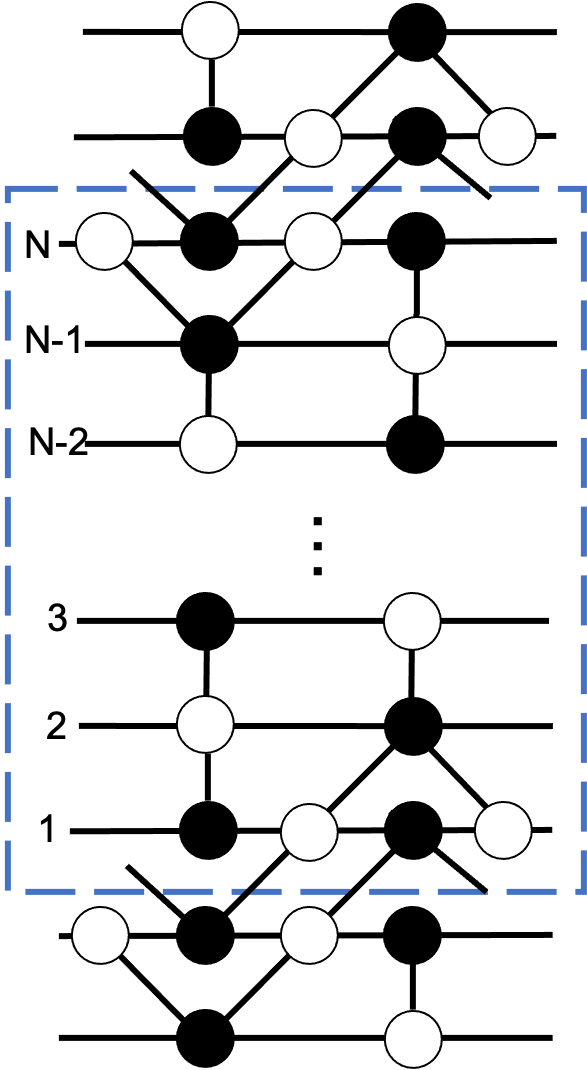}
    \caption{A proposed bipartite graphs to be folded to dimer for $D_N$ system with even $N$. The blue dashed square labels HALF unit cell in $U(2N)+8F$ dimer, which would become the dimer for $D_N$. 
    }
    \label{fig:D-dimer-new-1}
\end{figure}

We start with two double impurities introduced to the $A_{2N-4}$ dimer graph, placed as far apart from each other as possible. See Fig.~\ref{fig:D-dimer-new-1} for an illustration. 
The folding is performed in the following steps:
\begin{enumerate}
    \item Set $\tq_{2N-n+1}=-\tq_{n}$, $\tp_{2N-n+1} = -\tp_{2N-n+1}$, $n=1,\dots,N$. This gives $H_{1} = H_{2N-1}$. 
    \item Apply the change of canonical variables on the boundary of half-unit cell: 
    \[
        e^{\tq_1} \to \frac{\sinh\tq_1}{\cosh\frac{\tp_1}{2}}, \quad e^{\tp_1} \to \frac{\cosh\frac{\tp_1-2\tq_1}{2}}{\cosh\frac{\tp_1+2\tq_1}{2}}, \quad 
        e^{\tq_N} \to  \frac{\cosh \frac{\tp_N}{2} }{\sinh \tq_N}, \quad e^{\tp_N} \to \frac{\cosh\frac{\tp_N+2\tq_N}{2}}{\cosh \frac{\tp_N-2\tq_N}{2}}. 
    \]
    \item Modify the square loop as in \eqref{eq:mod-square}.
    \item The double impurity dimer now consists two identical copies of the half unit cell, with one rotated by 180 degrees.
\end{enumerate}

\begin{figure}
    \centering
    \includegraphics[width=0.6\textwidth]{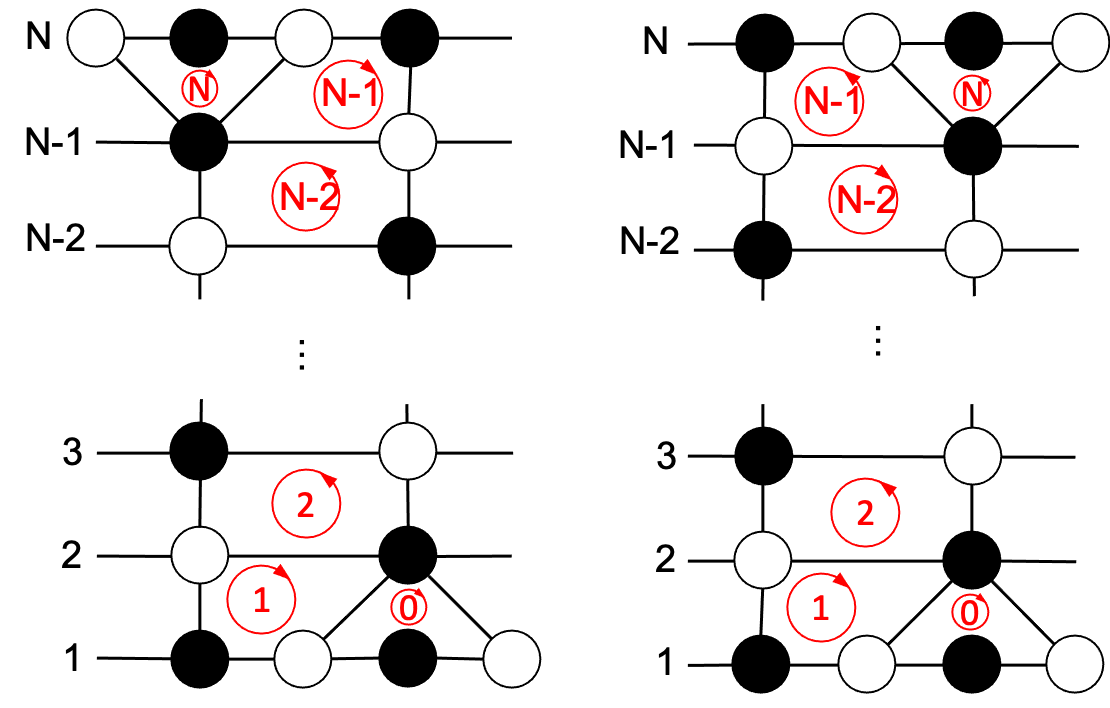}
    \caption{The dimer graph for $D_N$ theory, even $N$ on the left and odd $N$ on the right, with the fundamental square loops labeled.}
    \label{fig:DN-dimer}
\end{figure}

The  bipartite dimer diagram for $D_N$ model in Fig.~\ref{fig:DN-dimer} is based on the folding of the double impurity square dimer Fig.~\ref{fig:D-dimer-new-1}. All the 1-loops can be constructed by square loops and straight horizontal lines. 

The square loops obey the commutation relation
\begin{align}\label{eq:DN-commu}
\begin{split}
    & \{S_n,S_{n+1}\} = (-1)^n S_n S_{n+1}, \ n=0,\dots,N-1 \\
    & \{S_n,e^{\tp_n}\} = S_n e^{\tp_n}, \ \{e^{\tp_{n+1}},S_n\} = e^{\tp_{n+1}} S_n, \ n=1,\dots,N-1. 
\end{split}
\end{align}

The first Hamiltonian is the given by the sum over all 1-loops plus the two terms coming from connecting the two patchings: 
\begin{align}
\begin{split}
    H_1 = & \sum_{n=1}^N 2\cosh \tp_n + \sum_{n=1}^{N-1} S_n (e^{-(-1)^n\tp_n} + e^{-(-1)^{n+1}\tp_{n+1}}) \\
    & + S_1S_0 (e^{\tp_1}+e^{-\tp_2}) + S_{N-1}S_N (e^{-(-1)^{N-1}\tp_{N-1}}+e^{-(-1)^N\tp_N}) + e^{-2\tq_2} + e^{2\tq_{N-1}}
\end{split}
\end{align}
The square loops can be expressed in terms of canonical coordinates: 
\[
    S_n = e^{\tq_n-\tq_{n+1}} e^{(-1)^n\frac{\tp_n-\tp_{n+1}}{2}}, \quad n=1,\dots,N. \quad S_0 = e^{-2\tq_1}, \quad S_N = e^{2\tq_N}
\]
Thus we recover the first Hamiltonian of the $\hat{D}_N$ RTL:
\begin{align}
    H_1|_\text{dimer} = {\rm H}_{D_N}. 
\end{align}

\section{Lax formalism}\label{sec:Lax}


\subsection{Construct Lax matrix from dimer graph}

A Kasteleyn matrix $K_{\tb,\tw}(X,Y)$ encodes the structure of a given dimer graph. The mirror curve of the dimer graph is defined by
\begin{align}
    W(X,Y) = \det K_{\tb,\tw}(X,Y) = 0.
\end{align}
The mirror curve, when properly weighted the dimer graph's edges with respect to the canonical coordinates, coincides with the spectral curve of the integrable system
\begin{align}
    \tilde{W}(X,Y)=\det (\bT(X)-Y) = 0.
\end{align}
$\bT(X)$ is the monodromy matrix of the relativistic integrable system in Sklyanin's Lax formalism \cite{sklyanin1995separation}. 


The Kasteleyn matrix of $Y^{N,0}$ square dimer (Type A RTL) is a tri-diagonal $N\times N$ matrix given by
\begin{align}\label{eq:K-matrix-N0}
    K_{\tb_i,\tw_j} = \begin{cases} h_{i} - \tilde{h}{i} X, & \text{if } j = i \text{ and } i \in 2\BZ+1; \\ h{i} X^{-1} - \tilde{h}{i}, & \text{if } j = i \text{ and } i \in 2\BZ; \\ \tilde{v}{j} Y^{-\delta_{j,1}}, & \text{if } i = j - 1 \ (\text{mod } N); \\ v_{j} Y^{\delta_{j,N}}, & \text{if } i = j + 1 \ (\text{mod } N); \\ 0, & \text{otherwise}.
    \end{cases}
\end{align}
The 1-loops are constructed based on the dimer edges:
\begin{align}
\begin{split}
    d_j = h_{j} \tilde{h}_{j}^{-1} = e^{\tp_j}, \quad 
    c_j = h_{j} v_{j-1}^{-1} h_{j-1} \tilde{v}_j^{-1} = e^{\tq_j-\tq_{j+1}} e^{\frac{\tp_j+\tp_{j+1}}{2}}. 
\end{split}
\end{align}
To construct the $2\times2$ Lax matrices from the Kasteleyn matrix of the dimer graph, 
we consider the $N \times 1$ null vector of the Kastekeyn matrix
\[
    \psi = \begin{pmatrix}
        \psi_1 \\ \vdots \\ \psi_N
    \end{pmatrix}, \quad K_{\tb,\tw} \psi = 0. 
\]
Expanding the matrix equation gives $N$ equations
\begin{align}
\begin{split}
    & (h_j- \tilde{h}_j X^{-1}) \psi_j + v_{j-1} \psi_{j-1} + \tilde{v}_{j+1} \psi_{j+1} = 0, \ j \in 2\BZ+1; \\
    & (h_j X- \tilde{h}_j) \psi_j + v_{j-1} \psi_{j-1} + \tilde{v}_{j+1} \psi_{j+1} = 0, \ j \in 2\BZ. 
\end{split}
\end{align}
The vector components obeys the periodicity condition $\psi_{j+N} = Y \psi_j$. 
We shift
\[
    \psi_j \to 
    \begin{cases}
        \sqrt{X} \psi_j , &  j \in 2\BZ + 1 \\ 
        \frac{1}{\sqrt{X}} \psi_j , & n \in 2\BZ.
    \end{cases}
\]
to obtain $N$ equations 
\begin{align}\label{eq:frac-TQ}
    (X - h^{-1}_j \tilde{h}_j ) \psi_j + \sqrt{X} h^{-1}_j (v_{j-1}\psi_{j-1}+\tilde{v}_{j+1}\psi_{j+1}) = 0. 
\end{align}
Define $2\times 1$ vector
\begin{align}
    \Xi_j = \begin{pmatrix}
        \psi_j \\ \psi_{j-1}
    \end{pmatrix}.
\end{align}
The $N$ equations \eqref{eq:frac-TQ} can be organized into $N$ $2\times 2$ matrix equations
\begin{align}
    \sqrt{X} h_j^{-1} \tilde{v}_{j+1} \Xi_{j+1} = \begin{pmatrix}
        X - h_j^{-1} \tilde{h}_j & - \sqrt{X} h_j^{-1} v_{j-1} \\ 
        \sqrt{X} h_j^{-1} \tilde{v}_{j+1} & 0
    \end{pmatrix} \Xi_j = \sqrt{X} h_j^{-1} L_{j}(X) \Xi_{j}.
\end{align}
By assigning the edges with canonical coordinate, we obtain
\begin{align}
    L_j(X) = \begin{pmatrix}
        2\sinh \frac{x-\tp_j}{2} & - e^{-\tq_j} \\ e^{\tq_j} & 0
    \end{pmatrix}. 
\end{align}
The monodromy matrix is a ordered product of the Lax matrices
\begin{align}
    \bT(X) = L_N(X) \cdots L_1(X), \quad \bT(X) \Xi_{1}(X) = Y \Xi_1(X).  
\end{align}

\subsection{Reconstructing monodromy matrix from double impurity dimer}

\begin{figure}
    \centering
    \includegraphics[width=0.75\textwidth]{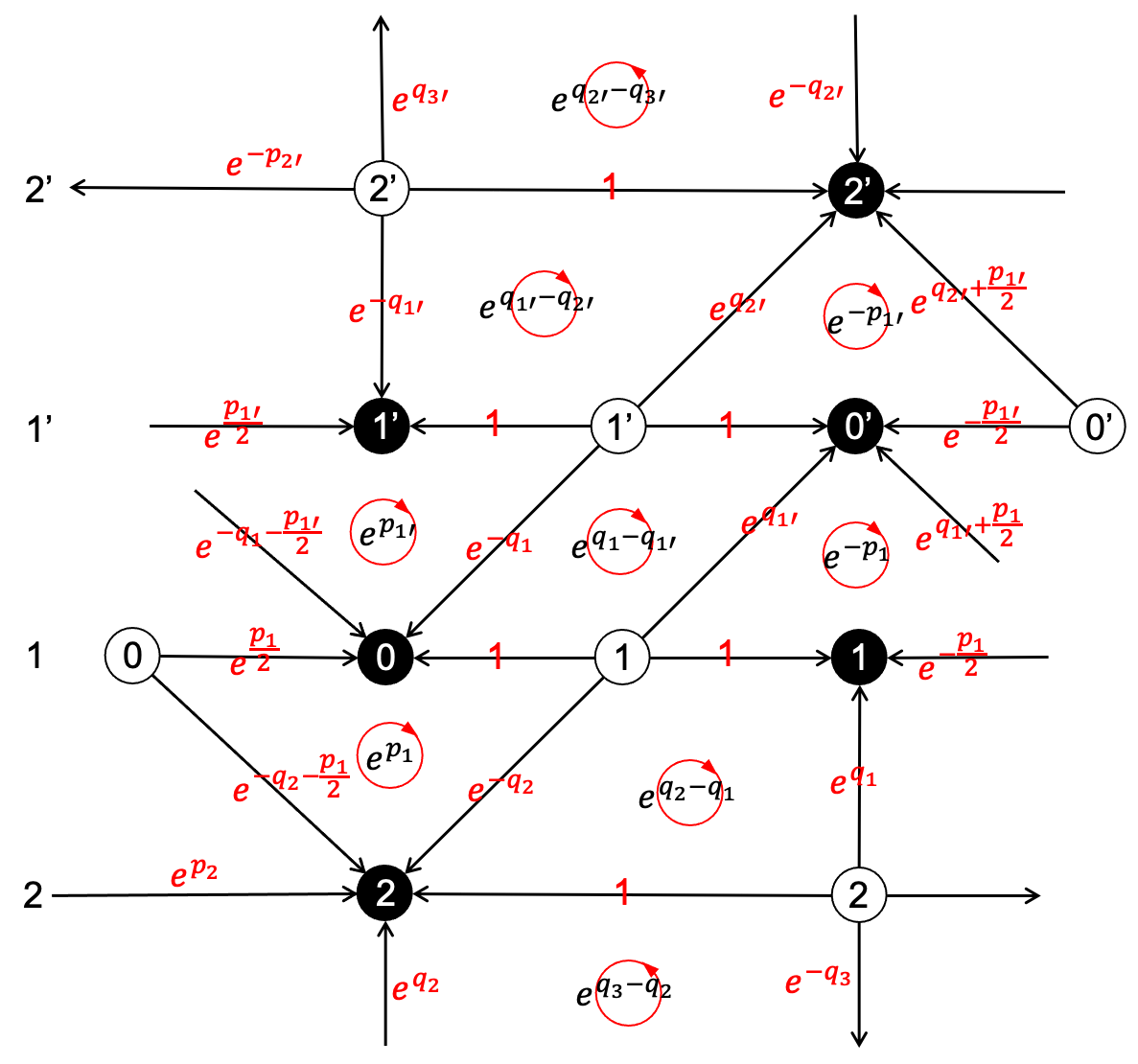}
    \caption{Assignment of the edges a weight associated to canonical coordinates which gives the correct loop contribution. Note that there exists more than one way to assign the edges. Different assignment are related by gauge transformation. }
    \label{fig:edges-double impurity}
\end{figure}

We consider the double impurity dimer of Fig.~\ref{fig:D-dimer-new-1}. Note that most of the structure of the type D dimer is the same as the type A square dimer, with the only difference being the introduction of the double impurity. See Fig.~\ref{fig:edges-double impurity} for an illustration. The submatrix of the full Kasteleyn matrix associated with the double impurity is

$K_{\tb,\tw}|_\text{DI}=$
\begin{tabular}{c|c c c c c c c c}
     & 3 & 2 & 1 & 0 & 0' & 1' & 2' & 3' \\
     \hline
    2 & $s_{23}$ & $h_{2}X^{-1}-\tilde{h}_{2}$ & $s_{21}$ & $s_{20}$ & 0 & 0 & 0 & 0 \\
    1 & 0 & $s_{12}$ & $h_1$ & $-{s_{10}}{X}$ & 0 & 0 & 0 & 0 \\
    0 & 0 & 0 & $-s_{01}$ & $h_0$ & $\frac{s_{00'}Y}{X}$ & $s_{01'}Y$ & 0 & 0 \\
    0' & 0 & 0 & $\frac{s_{0'1}}{Y}$ & $\frac{s_{0'0}X}{Y}$ & $-h_{0'}$ & $s_{0'1'}$ & 0 & 0 \\
    1' & 0 & 0 & 0 & 0 & $\frac{s_{1'0'}}{X}$ & $-h_{1'}$ & $s_{1'2'}$ & 0 \\
    2' & 0 & 0 & 0 & 0 & $s_{2'0'}$ & $s_{2'1'}$ & ${h_{2'}}-{\tilde{h}_{2'}}X$ & $s_{2'3'}$ \\
\end{tabular}

Note that $Y$ here labels the edge crosses HALF unit cell instead of the full unit cell. We would expect the spectral curve obtained from the Kasteleyn matrix as a function in $Y^2$. 

Consider the eigenvector $\psi=(\psi_N,\dots,\psi_0,\psi_{0'},\dots,\psi_{N'})$ of the Kasteleyn matrix.
\begin{subequations}\label{eq:Kasteleyn-BA}
\begin{align}
    & s_{23} \psi_3 + \left( \frac{h_2}{X} - {\tilde{h}_2} \right) \psi_2 + s_{21} \psi_1 + s_{20} \psi_0 = 0, \\
    & s_{12} \psi_2 + h_1 \psi_1 - {s_{10}}{X} \psi_0 = 0, \\
    & -s_{01} \psi_1 + h_0 \psi_0 + \frac{s_{00'}Y}{X} \psi_{0'} + s_{01'} Y \psi_{1'} = 0, \\
    & \frac{s_{0'1}}{Y} \psi_1 + \frac{s_{0'0}X}{Y} \psi_0 - h_{0'} \psi_{0'} + s_{0'1'} \psi_{1'} = 0, \\
    & \frac{s_{1'0'}}{X} \psi_{0'} - h_{1'} \psi_{1'} + s_{1'2'} \psi_{2'} = 0, \\
    & s_{2'0'} \psi_{0'} + s_{2'1'} \psi_{1'} + \left( {h_{2'}} - \tilde{h}_{2'} X \right) \psi_{2'} + s_{2'3'} \psi_{3'} = 0.
\end{align}
\end{subequations}
Combining the third and the fourth equations gives: 
\begin{subequations}
\begin{align}
    & \left( s_{01'}s_{0'1} + s_{0'1'}s_{01} \right) \psi_1 + \left( s_{0'1'}h_0 + s_{0'0}s_{01'} X \right) \psi_0 + \left( -s_{01'}h_{0'} + \frac{s_{00'}s_{0'1'}}{X} \right) Y \psi_{0'} = 0 \\
    & \left( s_{0'0}s_{01} X + s_{0'1}h_0 \right) \frac{1}{Y}\psi_0 + \left( - s_{01}h_{0'} + \frac{s_{00'}s_{0'1}}{X} \right) \psi_{0'} + \left( s_{0'1}s_{01'} + s_{01}s_{0'1'}  \right) \psi_{1'} = 0
\end{align}
\end{subequations}

We combine the matrices that transfers through the boundary by
\begin{align}
\begin{split}
    (1-e^{-2\tq_1}) \frac{Y}{X} (X-1)^2 \begin{pmatrix}
        \psi_{2} \\ \tilde{\psi}_1 = \psi_{1} + \frac{s_{20}}{s_{21}} \psi_0 
    \end{pmatrix} 
    & = \begin{pmatrix}
        TL & TR s_{21} \\ BL s_{21} & BR s_{21}^2
    \end{pmatrix} 
    \begin{pmatrix}
         \psi_{1'} + \frac{s_{2'0'}}{s_{2'1'}} \psi_{0'} \\
         \psi_{2'}
    \end{pmatrix} \\
    & = (1-e^{-2\tq_1}) \tilde{K}_1 \begin{pmatrix}
        \tilde\psi_{1'} = \psi_{1'} + \frac{s_{2'0'}}{s_{2'1'}} \psi_{0'} \\
        \psi_{2'}
    \end{pmatrix}
\end{split}
\end{align}

We assign the canonical coordinates to the edges across the double impurity according to Fig.~\ref{fig:edges-double impurity}. The four components of the reflection matrix $\tilde{K}_1$ are given by 
\[
    TL = - (1-e^{-2\tq_1}) \left[ X + \frac{1}{X} -  e^{-2\tq_1} (e^{\tp_1}-2+e^{-\tp_1}) - e^{\tp_1} - e^{-\tp_1} \right],
\]

\[
    TR = - (1-e^{-2\tq_1}) \left[ (e^{-\tp_1} + 1) (1-e^{-2\tq_1})  X - (e^{\tp_1}+1) (1-e^{-2\tq_1}) \right],
\]

\[
    BL = (1-e^{-2\tq_1}) \left[ (e^{-\tp_1} + 1) (1-e^{-2\tq_1})  \frac{1}{X} - (e^{\tp_1}+1) (1-e^{-2\tq_1}) \right],
\]

\[
    BR = (1-e^{-2\tq_1}) \left[ e^{-2\tq_1} \left( X+\frac{1}{X} \right) + (1-e^{-2\tq_1}) ( e^{-\tp_1} + e^{\tp_1} ) - 2 \right].
\]
See Appendix.~\ref{sec:detail-D} for computational detail. 
We further take canonical coordinate transformation \eqref{eq:change of variable} and modifies $s_{21}=s_{2'1'}$ according to \eqref{eq:mod-square}, which yields
\begin{align}
\begin{split}
    \tilde{K}_1 
    & =  \begin{pmatrix}
        X + \frac{1}{X} - 2 \cosh 2\tq_1 & 2\cosh\frac{\tp_1-2\tq_1}{2} X - 2\cosh \frac{\tp_1+2\tq_1}{2} \\
        -\frac{2}{X}\cosh \frac{\tp_1-2\tq_1}{2} + 2\cosh \frac{\tp_1+2\tq_1}{2} & - X - \frac{1}{X} + 2\cosh \tp_1
    \end{pmatrix}.
\end{split}
\end{align}
Similarly on the other side of the half unit cell, we obtain 
\begin{align}
\begin{split}
    \tilde{K}_N 
    & =  \begin{pmatrix}
        X + \frac{1}{X} - 2 \cosh 2\tq_N & 2\cosh\frac{\tp_1-2\tq_N}{2} X - 2\cosh \frac{\tp_N+2\tq_N}{2} \\
        -\frac{2}{X}\cosh \frac{\tp_N-2\tq_N}{2} + 2\cosh \frac{\tp_N+2\tq_N}{2} & - X - \frac{1}{X} + 2\cosh \tp_N
    \end{pmatrix}.
\end{split}
\end{align}

The monodromy matrix is defined by ordered product of the reflection matrices and Lax operators. 
With a proper gauge transformation, we obtain
\begin{align}
    \bT(X) = K_+(X) \tilde{L}_{2}(X) \cdots \tilde{L}_{N-1}(X) K_-(X) L_{N-1}(X) \cdots L_2(X).
\end{align}
The Lax matrices and reflection matrices are given by
\begin{subequations}
\begin{align}
    L_{j}(X) & = \begin{pmatrix}
        2\sinh \frac{x-\tp_j}{2} & -e^{-\tp_j} \\ e^{\tp_j} & 0
    \end{pmatrix}, \quad 
    \tilde{L}_j(X) = \begin{pmatrix}
        0 & e^{-\tq_{j}} \\ e^{\tq_{j}} & 2\sinh \frac{x+\tp_{j}}{2}
    \end{pmatrix}, \\
    K_+(X) & = \begin{pmatrix}
        2\sqrt{X}\cosh\frac{\tp_1-2\tq_1}{2}  - \frac{2}{\sqrt{X}}\cosh \frac{\tp_1+2\tq_1}{2} &  X + \frac{1}{X} - 2 \cosh 2\tq_1  \\
        - X - \frac{1}{X} + 2\cosh \tp_1  & 2\sqrt{X} \cosh \frac{\tp_1+2\tq_1}{2} -\frac{2}{\sqrt{X}}\cosh \frac{\tp_1-2\tq_1}{2} 
    \end{pmatrix}, \\
    K_-(X) & = \begin{pmatrix}
        2\sqrt{X} \cosh \frac{\tp_N+2\tq_N}{2} -\frac{2}{\sqrt{X}}\cosh \frac{\tp_N-2\tq_N}{2} & - X - \frac{1}{X} + 2\cosh \tp_N  \\
        X + \frac{1}{X} - 2 \cosh 2\tq_N &  2\sqrt{X}\cosh\frac{\tp_1-2\tq_N}{2}  - \frac{2}{\sqrt{X}}\cosh \frac{\tp_N+2\tq_N}{2}
    \end{pmatrix}.
\end{align}
\end{subequations}
The monodromy matrix obeys the eigenequation
\begin{align}
    \bT(X) 
    \begin{pmatrix}
        \psi_{2} \\ \tilde\psi_{1}
    \end{pmatrix} = Y^2 \left( {X} - \frac{1}{{X}} \right)^2 \begin{pmatrix}
        \psi_{2} \\ \tilde\psi_{1}
    \end{pmatrix}. 
\end{align}
The two components of the vector satisfies the spectral equation
\begin{align}
\begin{split}
    & Y^2 (X-X^{-1})^2 \left[ Y^2 (X-X^{-1})^2 - \Tr\bT(X) + \frac{(X-X^{-1})^2}{Y^2}  \right] \psi_2 = 0 \\
    & Y^2 (X-X^{-1})^2 \left[ Y^2 (X-X^{-1})^2 - \Tr\bT(X) + \frac{(X-X^{-1})^2}{Y^2}  \right] \tilde\psi_1 = 0
\end{split}
\end{align}
Recall that $Y$ labels the crossing of half unit cell. we recover the spectral curve in \eqref{eq:spectral-curve} after redefining $Y^2 \to Y$. 
In particular, the transfer matrix (trace of monodromy matrix) \eqref{def:generating} is the generating function of all conserving Hamiltonians.

\section{Summery and future directions} \label{sec:summary}

We construct the dimer graph for the type D relativistic Toda lattice by introducing a double impurity into the type A square dimer. The folding process from type A with impurity to the type D dimer involves identifying the faces and vertices in the dimer unit cell, followed by a change of variables as described in Eq.~\eqref{eq:change of variable} across the impurity insertion square. Additionally, certain square loops are modified by an overall factor as given in Eq.~\eqref{eq:mod-square}. 
The resulting Hamiltonian of the double impurity dimer indeed reproduces the $\hat{D}_N$ relativistic Toda lattice.

Furthermore, we construct the Lax matrix and reflection matrix from the Kasteleyn matrix of the new dimer graph using the Baker-Akhiezer function associated with the latter. By managing the coefficients of the Baker-Akhiezer function related to the double impurity and properly assigning the canonical coordinates to the edges of the dimer graph, we can reconstruct both reflection matrices at the boundaries.

Let us end this note with some potential future direction:
\begin{itemize}
    \item We are interested in the quantum type D relativistic Toda lattice (RTL). Recent studies show that the Bethe/Gauge correspondence provides an excellent tool, particularly for solving the wavefunction of quantum integrable systems. This correspondence has been explored for various non-relativistic integrable systems \cite{Nikita:V,Jeong:2023qdr,jeong2021intersecting,Chen:2019vvt,Chen:2020rxu,Lee:2020hfu,Jeong:2017pai,Jeong:2024hwf,Jeong:2024mxr}.

It is known that the Nekrasov-Shatashvili free energy, which works excellently in 4D, is insufficient for establishing the 5D Bethe/Gauge correspondence. Correct quantization requires including a tower of non-perturbative effects \cite{Grassi:2014zfa,Franco:2015rnr,Hatsuda:2015qzx}, which can be addressed by introducing a Wilson-loop/quantum mirror map via topological string theory \cite{Grassi:2017qee,Grassi:2014zfa}. Off-shell quantization for type A RTL has been explored in \cite{Lee:2023wbf}.
    \item Find potential dimer graphs for type B, C, and E relativistic Toda lattices (RTLs). We anticipate that this can be achieved by appropriately introducing impurities into the embedded type A dimer graph, followed by a folding process.

    \item A Poisson graph (or quiver graph) can be constructed based on the blowup of the Lie algebra Dynkin diagram. Studying the quiver graph provides insights into the spectral properties of the relativistic Toda lattice (RTL) \cite{Kruglinskaya:2014pza}. In principle, the dimer graph associated with the same RTL should be the dual graph of the Poisson graph. This relation is straightforward in the case of type A.

    For types B, C, D, and E, the situation is more complex, as the dual of the quiver graph derived from the Dynkin diagram is often non-planar. It is an interesting topic to explore the potential relationship between the dimer graph we constructed for type D and the Poisson graph.
    \item In this note, we introduced a double impurity placed as far apart as possible, based on the toric diagram shared by $A_{2N-1}/8F$ and $D_{N}$. As noted, this is not the only way to modify the dimer graph to construct different but potentially dual integrable systems. Similar observations have been made in \cite{Lee:2023wbf}. 
\end{itemize}

\newpage 
\appendix

\section{Lax formalism} \label{sec:type A dimer}
Let us briefly illustrate how the type A periodic relativistic Toda lattice (RTL) is associated with the $Y^{N,0}$ square dimer. We will also demonstrate how the $N \times N$ Lax matrix can be derived from the dimer Kasteleyn matrix.

\begin{figure}[h]
    \centering
    \includegraphics[width=0.5\textwidth]{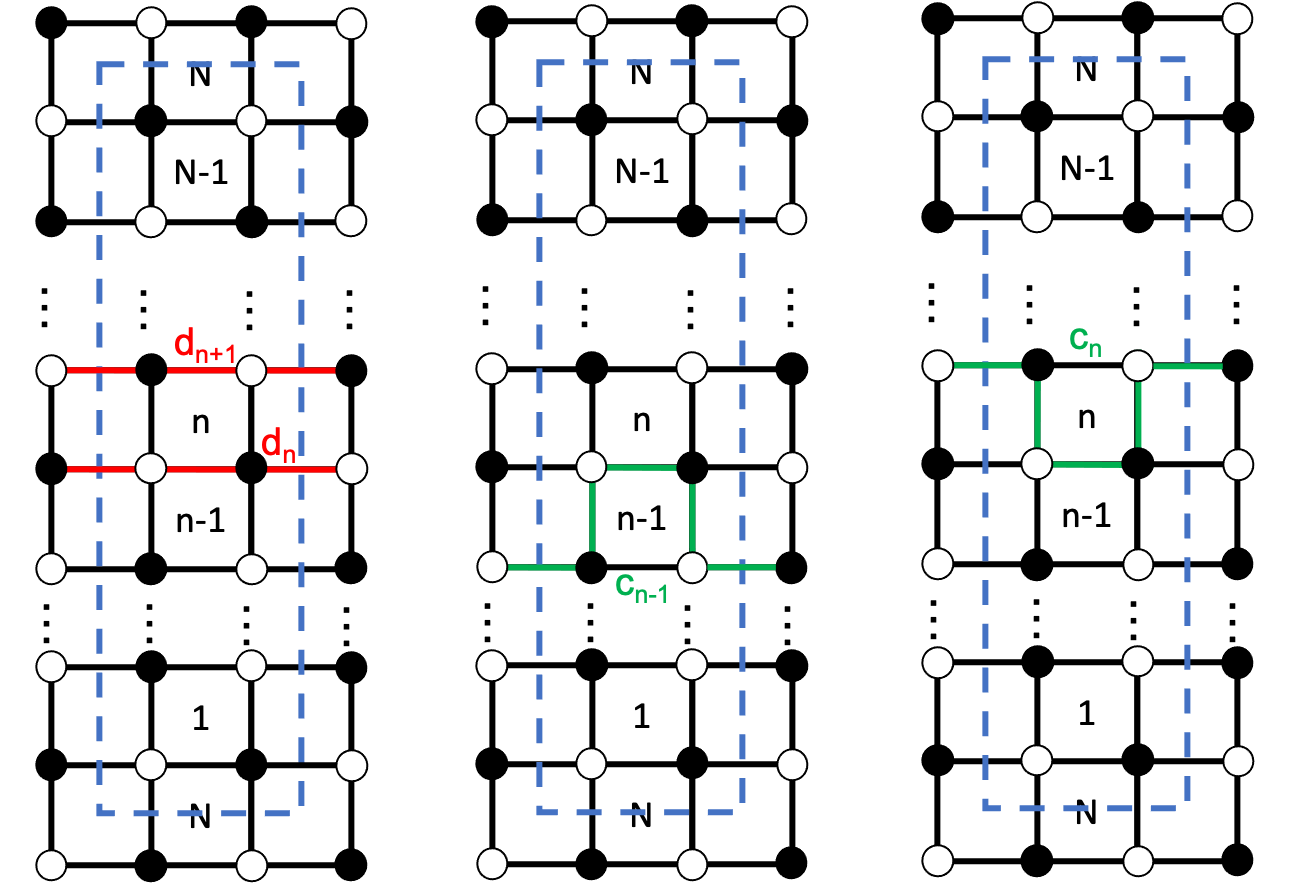}
    \caption{The brane tiling for $Y^{N,0}$ dimer model with $N$ even. A unit cell is encircled by the dashed blue line. This dimer is associated to the $\hat{A}_{N-1}$ relativistic Toda lattice.}
    \label{fig:YN0-loop}
\end{figure}

We choose the reference perfect matching on the $Y^{N,0}$ square dimer, as shown in Fig.~\ref{fig:YN0}, to be the horizontal edges with white nodes on the left and black nodes on the right. There are $2N$ 1-loops, denoted $c_n$ and $d_n$ for $n=1, \dots, N$, which satisfy the commutation relations based on their shared edges:
\begin{align}\label{eq:c-d commu}
    \{c_n,d_n\} = c_n d_n, \quad \{d_{n+1},c_n\} = c_nd_{n+1}, \quad \{c_n,c_{n+1}\} = -c_n c_{n+1}. 
\end{align}
with the periodicity $c_{N+1}=c_1$, $d_{N+1}=d_1$. 
See Fig.~\ref{fig:YN0-loop} for illustration.

The 1-loops can be written in terms of the canonical coordinates by
\begin{align}
    d_n = e^{\tp_n}, \ c_n = e^{\tq_n - \tq_{n+1}} e^{\frac{\tp_n+\tp_{n+1}}{2}}, \ n=1,\dots,N. 
\end{align}
The canonical coordinates obeys the periodicity $\tq_{N+1}=\tq_1$, $\tp_{N+1}=\tp_1$. 
The first Hamiltonian of the $Y^{N,0}$ dimer graph recovers the $\hat{A}_{N-1}$ RTL Hamiltonian
\begin{align}\label{eq:RTL-H}
    H_1|_{Y^{N,0}} = \sum_{n=1}^N c_n + d_n = \sum_{n=1}^N e^{\tp_n} + {g}^2 e^{\tq_n - \tq_{n+1}} e^{\frac{\tp_n+\tp_{n+1}}{2}} = {\rm H}_{\hat{A}_{N-1}}
\end{align}

\subsection{$N\times N$ Lax matrix}
There are two ways to construct the Lax matrix of the integrable system from the dimer graph. The first one is by studying the equation of motion of the 1-loops, the second through Kasteleyn matrix of the dimer graph.

\paragraph{Equation of Motion: }
The equations of motion of the coordinates $\{c_n,d_n\}$ given by the Haamiltonian \eqref{eq:RTL-H} have the following form
\begin{align}\label{eq:time-1-loop}
\begin{split}
    \dot{c}_n & = \{c_n,H_1\} =  c_n(d_n-d_{n+1} + c_{n-1}-c_{n+1}), \\
    \dot{d}_n & = \{d_n,H_1\} = d_n(c_{n-1}-c_{n}).
\end{split}
\end{align}
and can be write as compatibility condition for the two linear problems \cite{Kuznetsov:1994ur,bruschi1989lax}
\begin{align}
\begin{split}
    & (X - d_j) \phi_j + \sqrt{X} (f_{j}\phi_{j+1} + f_{j-1}\phi_{j-1}) = 0 \\
    & \dot{\phi}_j = -\frac{1}{2} (c_j-c_{j-1}+X) \phi_j - \frac{\sqrt{X}}{2} (f_j \phi_{j+1}-f_{j-1}\phi_{j-1})
\end{split}
\end{align}
It can be cast into matrix form 
\begin{align}
    L \boldsymbol\phi = 0, \ \dot{\boldsymbol\phi} = -M \boldsymbol\phi
\end{align}
The Hamiltonian system obeys a "weak" Lax triad 
\begin{align}
    \frac{d}{dt}{L} = [L,M] + C L
\end{align}
with the matrices
\begin{subequations}
\begin{align}
    L_{ij} & = (X - d_i) \d_{i,j} + \sqrt{X} f_i \d_{i+1,j} + \sqrt{X} f_{i-1} \d_{i-1,j} \\
    M_{ij} & = \frac{1}{2} \left[ (c_i-c_{i-1}+X) \d_{i,j} + \sqrt{X} f_i \d_{i+1,j} - \sqrt{X} f_{i-1} \d_{i-1,j} \right] \\
    C_{ij} & = (c_i-c_{i-1})\d_{i,j}
\end{align}
\end{subequations}
with $f_i^2 = c_i$. 
The matrix $C$ is diagonal and traceless. This leads to 
\begin{align}
\begin{split}
    \Tr L^{-1} \frac{dL}{dt} 
    = \Tr L^{-1} \left(  [L,M] + C L \right) 
    = \Tr C = 0. 
\end{split}
\end{align}
By using Abel identity 
\[
    \frac{d}{dt} \log \det L = \Tr L^{-1} \frac{dL}{dt} = 0
\]
one concludes $\det L(X,Y)$ is the generating function of the integral of motion of RTL. 

\paragraph{Kasteleyn matrix: }
The Kasteleyn matrix of square dimer in Fig.\ref{fig:YN0-loop} is
\begin{align}
    K_{\tb_n,\tw_m} = \begin{cases}
        (h_{n}-\tilde{h}_{n} X^{-1}), & n=m \in 2\BZ+1; \\
        (h_{n}X-\tilde{h}_{n}), & n=m \in 2\BZ;
        \\
        \tilde{v}_{m} Y^{\d_{m,1}} , & n=m-1 \ \text{mod}(N); \\
        v_{m} Y^{-\d_{m,N}} & n=m+1, \ \text{mod}(N); \\
        0, & \text{otherwise}
    \end{cases}
\end{align}
The 1-loops are constructed through the dimer edges
\begin{align}
\begin{split}
    d_n & = h_{n}^{-1} \tilde{h}_{n} = e^{\tp_n}, \quad c_n = h_{n}^{-1} v_{n-1} h_{n-1}^{-1} \tilde{v}_n = e^{\tq_n-\tq_{n+1}} e^{\frac{\tp_n+\tp_{n+1}}{2}}. 
\end{split}
\end{align}
We consider $N \times 1$ vector
\[
    \boldsymbol\psi = \begin{pmatrix}
        \psi_1 \\ \psi_2 \\ \vdots \\ \psi_N
    \end{pmatrix}, \quad K_{\tb,\tw} \boldsymbol\psi = 0
\]
Expanding the matrix equation gives $N$ equations
\begin{align}
\begin{split}
    & (h_n- \tilde{h}_n X^{-1}) \psi_n + v_{n-1} \psi_{n-1} + \tilde{v}_{n+1} \psi_{n+1} = 0, \ n \in 2\BZ+1; \\
    & (h_n X- \tilde{h}_n) \psi_n + v_{n-1} \psi_{n-1} + \tilde{v}_{n+1} \psi_{n+1} = 0, \ n \in 2\BZ. 
\end{split}
\end{align}
with periodicity $\psi_{n+N} = y \psi_n$. 
By shifting
\[
    \psi_n \to 
    \begin{cases}
        \sqrt{X} \psi_n , &  n \in 2\BZ + 1 \\ 
        \frac{1}{\sqrt{X}} \psi_n , & n \in 2\BZ.
    \end{cases}
\]
We obtain $N$ equations 
\begin{align}
    (X - h^{-1}_n \tilde{h}_n ) \psi_n + \sqrt{X} h^{-1}_n (v_{n-1}\psi_{n-1}+\tilde{v}_{n+1}\psi_{n+1}) = 0
\end{align}
which can be organized into a matrix equation
\begin{align}
    \tilde{L} \boldsymbol\psi = 0, \ \tilde{L}_{nm} = (X - h^{-1}_n \tilde{h}_n ) \d_{n,m} + \sqrt{X} h^{-1}_n (v_{n-1} \d_{n-1,m} +\tilde{v}_{n+1} \d_{n+1,m})
\end{align}
We further conjugate diagonal matrix
$A = \text{diag} (a_1,\dots,a_N)$ with
\[
    \frac{a_n}{a_{n-1}} = \sqrt{\frac{\tilde{v}_n h_n}{v_{n-1}h_{n-1}}}
\]
to obtain
\begin{align}
    L_{nm} =(A\tilde{L}A^{-1})_{nm} = (X - d_n ) \d_{nm} + \sqrt{X} ( f_n \d_{n-1,m} + f_{n+1} \d_{n+1,m})
\end{align}
with $f_n^2 = c_n$.

\subsection{$2\times 2$ Lax matrix}
The $2\times 2$ Lax formalism proposed by Sklyanin is equivalent to the $N \times N$ Lax matrix. 
Define $2 \times 1$ vector
\begin{align}
    \Xi_n = \begin{pmatrix}
        \psi_n \\ \psi_{n-1}
    \end{pmatrix}
\end{align}
The $N$ equations \eqref{eq:frac-TQ} can be organized into $N$ $2\times 2$ matrix equation
\begin{align}
    \sqrt{X} h_n^{-1} \tilde{v}_{n+1} \Xi_{n+1} = \begin{pmatrix}
        X - h_n^{-1} \tilde{h}_n & - \sqrt{X} h_n^{-1} v_{n-1} \\ 
        \sqrt{X} h_n^{-1} \tilde{v}_{n+1} & 0
    \end{pmatrix} \Xi_n = \sqrt{X} h_n^{-1} L_{n}(X) \Xi_{n}
\end{align}
The monodromy matrix is defined by 
\begin{align}
    \bT(X) = L_N(x) \cdots L_1(X). 
\end{align}

\section{Canonical change of variables}\label{sec:change-of-variables}
Let us check the change of variables in \eqref{eq:change of variable} is canonical by direct computation: 
\begin{align}
\begin{split}
    e^{\tq} e^{\tp} & = \{ e^{\tq} , e^{\tp} \} \\
    & \to \left\{ \frac{\cosh\frac{\tp}{2}}{\sinh\tq}, \frac{\cosh\frac{\tp-2\tq}{2}}{\cosh\frac{\tp+2\tq}{2}} \right\} \\
    & = \frac{\p}{\p\tq} \frac{\cosh\frac{\tp}{2}}{\sinh\tq} \frac{\p}{\p\tp} \frac{\cosh\frac{\tp-2\tq}{2}}{\cosh\frac{\tp+2\tq}{2}} - \frac{\p}{\p\tq} \frac{\cosh\frac{\tp-2\tq}{2}}{\cosh\frac{\tp+2\tq}{2}} \frac{\p}{\p\tp} \frac{\cosh\frac{\tp}{2}}{\sinh\tq} \\
    & = -\frac{\cosh\frac{\tp}{2}\cosh\tq}{\sinh^2\tq} \frac{1}{2} \frac{\sinh\frac{\tp-2\tq}{2}\cosh\frac{\tp+2\tq}{2} - \cosh\frac{\tp-2\tq}{2}\sinh\frac{\tp+2\tq}{2}}{\cosh^2\frac{\tp+2\tq}{2}} \\
    & \qquad + \frac{\sinh\frac{\tp-2\tq}{2}\cosh\frac{\tp+2\tq}{2}+\cosh\frac{\tp-2\tq}{2}\sinh\frac{\tp+2\tq}{2}}{\cosh^2\frac{\tp+2\tq}{2}} \frac{1}{2} \frac{\sinh\frac{\tp}{2}}{\sinh\tq} \\
    & = \frac{1}{2} \frac{\sinh\tp\sinh\frac{\tp}{2}\sinh\tq + \cosh\frac{\tp}{2}\cosh\tq \sinh2\tq}{\sinh^2\tq \cosh^2\frac{\tp+2\tq}{2}} \\
    & = \frac{\cosh\frac{\tp}{2}\sinh^2\frac{\tp}{2} + \cosh\frac{\tp}{2}\cosh^2\tq }{\sinh\tq \cosh^2\frac{\tp+2\tq}{2}} \\
    & = \frac{\cosh\frac{\tp}{2}}{\sinh\tq} \frac{\cosh\frac{\tp-2\tq}{2}}{\cosh\frac{\tp+2\tq}{2}}
\end{split}
\end{align}

\section{Detail computation in section 4.2}\label{sec:detail-D}

The six components of the Kasteleyn matrix Baker-Akzeihier function \eqref{eq:Kasteleyn-BA} across the double impurity insertion can be organized into six matrix equations
\begin{subequations}
\begin{align}
    & s_{23} \begin{pmatrix}
        \psi_{3} \\ \psi_{2}
    \end{pmatrix} = \begin{pmatrix}
        \tilde{h}_2 - \frac{h_2}{X} & -s_{21} \\ s_{23} & 0
    \end{pmatrix} \begin{pmatrix}
        \psi_{2} \\ \psi_{1} + \frac{s_{20}}{s_{21}} \psi_{0}
    \end{pmatrix} \\
    & s_{12} \begin{pmatrix}
        \psi_{2} \\ \psi_{1} + \frac{s_{20}}{s_{21}} \psi_0
    \end{pmatrix} = \begin{pmatrix}
        -h_1 & {s_{10}}{X} \\ s_{12} & s_{12} \frac{s_{20}}{s_{21}}
    \end{pmatrix} \begin{pmatrix}
        \psi_{1} \\ \psi_{0}
    \end{pmatrix} \\
    & (-s_{0'1'}s_{01}-s_{01'}s_{0'1}) \begin{pmatrix}
        \psi_{1} \\ \psi_{0}
    \end{pmatrix} = \begin{pmatrix}
        s_{0'1'}h_0 + s_{0'0}s_{01'} X &  h_{0'}s_{01'} + \frac{s_{00'}s_{0'1'}}{X} \\ -s_{0'1'}s_{01} - s_{01'}s_{0'1} & 0
    \end{pmatrix} \begin{pmatrix}
        \psi_{0} \\ Y \psi_{0'}
    \end{pmatrix} \\
    & \frac{1}{Y} (s_{01}s_{0'0}X+h_0s_{0'1}) \begin{pmatrix}
        \psi_{0} \\ Y \psi_{0'}
    \end{pmatrix} = \begin{pmatrix}
        h_{0'}s_{01} - \frac{s_{00'}s_{0'1}}{X} & -s_{01}s_{0'1'}-s_{0'1}s_{01'} \\ s_{01}s_{0'0}X + h_0 s_{0'1} & 0
    \end{pmatrix} \begin{pmatrix}
        \psi_{0'} \\ \psi_{1'}
    \end{pmatrix} \\
    & (s_{2'0'}h_{1'}X-s_{1'0'}s_{2'1'}) \begin{pmatrix}
        \psi_{0'} \\ \psi_{1'}
    \end{pmatrix} = X s_{2'1'} \begin{pmatrix}
        -h_{1'} & s_{1'2'} \\ -\frac{s_{1'0'}}{X} & - s_{1'2'}\frac{s_{2'0'}}{s_{2'1'}}
    \end{pmatrix} \begin{pmatrix}
        \psi_{1'} + \frac{s_{2'0'}}{s_{2'1'}} \psi_{0'} \\ \psi_{2'}
    \end{pmatrix} \\
    & s_{2'1'} \begin{pmatrix}
        \psi_{1'} + \frac{s_{2'0'}}{s_{2'1'}} \psi_{0'} \\ \psi_{2'}
    \end{pmatrix} = \begin{pmatrix}
        {\tilde{h}_{2'}}{X} - h_{2'} & -s_{2'3'} \\ s_{2'1'} & 0
    \end{pmatrix} \begin{pmatrix}
        \psi_{2'} \\ \psi_{3'}
    \end{pmatrix}
\end{align}
\end{subequations}
We will combine the matrices in the second to fifth line
\begin{align}
\begin{split}
     & \begin{pmatrix}
        -h_1 & {s_{10}}{X} \\ s_{12} & s_{12} \frac{s_{20}}{s_{21}}
    \end{pmatrix}  \begin{pmatrix}
        s_{0'1'}h_0 + s_{0'0}s_{01'} X &  h_{0'}s_{01'} + \frac{s_{00'}s_{0'1'}}{X} \\ -s_{0'1'}s_{01} - s_{01'}s_{0'1} & 0
    \end{pmatrix} \\
    & \times \begin{pmatrix}
        h_{0'}s_{01} - \frac{s_{00'}s_{0'1}}{X} & -s_{01}s_{0'1'}-s_{0'1}s_{01'} \\ s_{01}s_{0'0}X + h_0 s_{0'1} & 0
    \end{pmatrix} \begin{pmatrix}
        -h_{1'} & s_{1'2'} \\ -\frac{s_{1'0'}}{X} & - s_{1'2'}\frac{s_{2'0'}}{s_{2'1'}}
    \end{pmatrix} \\
    = & \begin{pmatrix}
        1 & 0 \\ 0 & s_{12}
    \end{pmatrix} 
    \begin{pmatrix}
        - [s_{10} (s_{01}s_{0'1'}+s_{01'}s_{0'1}) + h_1s_{0'0}s_{01'}]X - h_1h_0s_{0'1'} & -h_1h_{0'}s_{01'} - \frac{h_1s_{00'}s_{0'1'}}{X} \\
        s_{0'0}s_{01'}X + s_{0'1'}h_0 - (s_{01}s_{0'1'}+s_{01'}s_{0'1}) \frac{s_{20}}{s_{21}} & h_{0'} s_{01'} + \frac{s_{00'}s_{0'1'}}{X}
    \end{pmatrix} \\
    & \times 
    \begin{pmatrix}
        -h_0h_1s_{01} + \frac{(s_{01}s_{0'1'}+s_{01'}s_{0'1})s_{1'0'}+h_{1'}s_{00'}s_{0'1}}{X} & \left[ h_0s_{01} + (s_{01}s_{0'1'}+s_{01'}s_{0'1})\frac{s_{20}}{s_{21}} \right] - \frac{s_{00'}s_{01'}}{X} \\
        -h_{1'} s_{01} s_{0'0} X - h_{1'}h_0 s_{0'1} & s_{01}s_{0'0} X + h_0 s_{0'1}
    \end{pmatrix} \begin{pmatrix}
        1 & 0 \\ 0 & s_{12}
    \end{pmatrix} \\
    = & \begin{pmatrix}
        TL & TR s_{12} \\ BL s_{12} & BR s_{12}^2
    \end{pmatrix} 
\end{split}
\end{align}

The folding restricts: 
\[
    h_0 = h_{0'}, \ h_1 = h_{1'}, \ s_{01} = s_{0'1'}, \ s_{10} = s_{1'0'}, \ s_{12} = s_{1'2'}, \ s_{01'}=s_{0'1}, \ s_{0'0}=s_{00'}.
\]
\begin{align}
\begin{split}
    & \begin{pmatrix}
        1 & 0 \\ 0 & s_{12}
    \end{pmatrix} 
    \begin{pmatrix}
        [s_{10} (s_{01}s_{0'1'}-s_{01'}s_{0'1}) - h_1s_{0'0}s_{01'}]X - h_1h_0s_{0'1'} & -h_1h_{0'}s_{01'} - \frac{h_1s_{00'}s_{0'1'}}{X} \\
        s_{0'0}s_{01'}X + s_{0'1'}h_0 + (s_{01}s_{0'1'}-s_{01'}s_{0'1}) \frac{s_{20}}{s_{21}} & h_{0'} s_{01'} + \frac{s_{00'}s_{01}}{X}
    \end{pmatrix} \\
    & \times 
    \begin{pmatrix}
        -h_0h_1s_{01} + \frac{(s_{01}^2-s_{01'}s_{0'1})s_{10}-h_1s_{00'}s_{0'1}}{X} & -\left[ h_0s_{01} + (s_{01}^2-s_{01'}s_{0'1})\frac{s_{20}}{s_{21}} \right] - \frac{s_{00'}s_{01'}}{X} \\
        h_1 s_{01} s_{0'0} X + h_1h_0 s_{0'1} & s_{01}s_{0'0} X + h_0 s_{0'1}
    \end{pmatrix} \begin{pmatrix}
        1 & 0 \\ 0 & s_{12}
    \end{pmatrix} \\
    & = \begin{pmatrix}
        TL & TR s_{12} \\ BL s_{12} & BR s_{12}^2
    \end{pmatrix} 
\end{split}
\end{align}
\begin{align}
\begin{split}
    TL = & - ( s_{01}^2 - s_{01'}s_{0'1} ) h_0h_1s_{01}s_{10}  \left( X + \frac{1}{X} \right) \\
    & + ( s_{10}s_{01}^2 - s_{10} s_{01'}s_{0'1} - h_1 s_{0'0}s_{01'} )^2 + h_0^2h_1^2(s_{01}^2-s_{0'1}s_{01'}) - h_1^2 s_{01}^2 s_{00'}s_{0'0} \\
    = & (s_{01}^2-s_{01'}^2)  \left[ - s_{01}s_{10} h_0 h_1 \left( X + \frac{1}{X} \right) + s_{10}^2(s_{01}^2-s_{01'}^2) - 2 {s_{0'0}s_{01'}}{h_1s_{10}} +{h_0^2h_1^2} - h_1^2 s_{00'}^2  \right] 
\end{split}
\end{align}
\begin{align}
\begin{split}
    TR = & \left[- ( s_{10}s_{01}^2 - s_{10} s_{01'}s_{0'1} - h_1 s_{0'0}s_{01'} ) \left( h_0s_{01} + (s_{01}^2-s_{01'}s_{0'1}) \frac{s_{20}}{s_{21}} \right)   - h_1h_0s_{01'}s_{01}s_{0'0} \right] X \\ 
    & + \left[ h_1h_0s_{01}s_{00'}s_{01'} - h_1h_0s_{00'}s_{01}s_{01'} \right] \frac{1}{X} \\
    & - ( s_{10}(s_{01}^2 - s_{01'}s_{0'1}) - h_1 s_{0'0}s_{01'} ) s_{00'}s_{01'} + h_1h_0s_{01} \left( h_0s_{01} + (s_{01}^2-s_{01'}s_{0'1}) \frac{s_{20}}{s_{21}} \right) \\
    & - h_1h_0^2 s_{01'}s_{0'1} - h_1s_{00'}s_{0'0}s_{01}^2 \\
    = & (s_{01}^2-s_{0'1}^2) \left[ -\frac{s_{10}s_{20}}{s_{21}}(s_{01}^2-s_{0'1}^2) - h_{0}s_{01}s_{10} + h_1s_{0'0}s_{01'} \frac{s_{20}}{s_{21}}  \right] X  \\
    & + (s_{01}^2-s_{0'1}^2) \left[ h_1h_0^2 - h_1 s_{00'}s_{0'0} - s_{10}s_{00'}s_{01'} + h_1h_0s_{01} \frac{s_{20}}{s_{21}} \right]  
\end{split}
\end{align}
\begin{align}
\begin{split}
    BL = & [-h_1h_0s_{01}s_{0'0}s_{01'}+h_1h_0s_{0'1}s_{0'0}s_{01}] X \\
    & + \left[ ( s_{10}(s_{01}^2 - s_{01'}s_{0'1}) - h_1 s_{00'}s_{0'1} ) \left( h_0s_{01} + (s_{01}^2-s_{01'}s_{0'1}) \right) \frac{s_{20}}{s_{21}}  + h_1h_0s_{01'}s_{01}s_{0'0} \right] \frac{1}{X} \\
    & + ( (s_{01}^2-s_{01'}s_{0'1})s_{10} - h_1 s_{00'}s_{0'1} ) s_{0'0}s_{01'} - h_0h_1s_{01} \left( h_0s_{01} + (s_{01}^2-s_{01'}s_{0'1}) \frac{s_{20}}{s_{21}} \right) \\
    & + h_1s_{01}^2 s_{00'}s_{0'0} + h_0^2h_1 s_{0'1}s_{01'} \\
    = & (s_{01}^2-s_{01'}^2) \left[ \frac{s_{10}s_{20}}{s_{21}}(s_{01}^2-s_{0'1}^2) + h_0s_{01}s_{10} - h_{1}s_{00'}s_{0'1} \frac{s_{20}}{s_{21}} \right] \frac{1}{X} \\
    & + (s_{01}^2-s_{01'}^2) \left[ -h_1h_0^2 + h_1 s_{00'}s_{0'0} + s_{10}s_{00'}s_{01'} - h_1h_0s_{01} \frac{s_{20}}{s_{21}} \right]
\end{split}
\end{align}
\begin{align}
\begin{split}
     BR = & - s_{0'0}s_{01'} (s_{01}^2-s_{01'}s_{0'1}) \frac{s_{20}}{s_{21}} \left( X + \frac{1}{X} \right) \\
    & - \left( h_0s_{01} + (s_{01}^2-s_{0'1}s_{01'}) \frac{s_{20}}{s_{21}} \right)^2 - s_{0'0}s_{01'}s_{00'}s_{01'} + h_0^2 s_{01'}s_{0'1} + s_{00'}s_{0'0} s_{01}^2 \nonumber \\
    = & (s_{01}^2-s_{01}^2) \left[ - s_{0'0}s_{01'}\frac{s_{20}}{s_{21}} \left( X + \frac{1}{X} \right) + (s_{01}^2-s_{01'}^2) \frac{s_{20}^2}{s_{21}^2} - 2 h_0s_{01} \frac{s_{20}}{s_{21}} + s_{00'}s_{0'0} - h_0^2 \right] 
\end{split}
\end{align}
We assign the edges on the double impurity according to Fig.~\ref{fig:edges-double impurity}. 
 We obtain:
\[
    TL = - (1-e^{-2\tq_1}) \left[ X + \frac{1}{X} -  e^{-2\tq_1} (e^{\tp_1}-2+e^{-\tp_1}) - e^{\tp_1} - e^{-\tp_1} \right]
\]

\[
    TR = - (1-e^{-2\tq_1}) \left[ (e^{-\tp_1} + 1) (1-e^{-2\tq_1})  X - (e^{\tp_1}+1) (1-e^{-2\tq_1}) \right]
\]

\[
    BL = (1-e^{-2\tq_1}) \left[ (e^{-\tp_1} + 1) (1-e^{-2\tq_1})  \frac{1}{X} - (e^{\tp_1}+1) (1-e^{-2\tq_1}) \right]
\]

\[
    BR = (1-e^{-2\tq_1}) \left[ e^{-2\tq_1} \left( X+\frac{1}{X} \right) + (1-e^{-2\tq_1}) ( e^{-\tp_1} + e^{\tp_1} ) - 2 \right]
\]

\newpage
\bibliographystyle{utphys}
\bibliography{SO-SU}

\end{document}